\documentclass[3p,final,11pt,authoryear]{elsarticle}
\usepackage[colorlinks]{hyperref}
\usepackage[utf8]{inputenc}
\usepackage{graphicx}
\usepackage{mathtools}
\usepackage{amsmath}
\usepackage{lscape}
\usepackage[nameinlink,noabbrev]{cleveref}
\usepackage{enumitem}
\usepackage{natbib}
\usepackage{subcaption}
\usepackage[tableposition = top]{caption}
\usepackage{booktabs}
\usepackage{algorithm}
\usepackage{algorithmic}
\usepackage{hyperref}
\usepackage{float}
\usepackage{lineno}
\usepackage{tabularx}
\usepackage{adjustbox}
\usepackage{rotating}
\usepackage{lscape}
\usepackage{amsfonts}
\usepackage{placeins}
\usepackage{tabularx}
\usepackage{multirow}
\usepackage{pdflscape}
\usepackage{booktabs}
\usepackage{adjustbox}
\usepackage{subcaption}
\usepackage{threeparttable}
\usepackage{makecell}
\usepackage{longtable}

\setcitestyle{authoryear,open={(},close={)}}
\usepackage{setspace}

\begin{document}
\begin{frontmatter}

\title{Exclusive and Shared Electric Flying Taxis: Evidence on Modal Shares, Stated Reasons, and Modal Shifts} 

    \author[1]{Nael Alsaleh\corref{cor1}}\ead{nael.alsaleh@aurak.ac.ae}
    \author[2]{Tareq Alsaleh}\ead{talsaleh@torontomu.ca}
    \author[1]{Fayez Moutassem}\ead{fayez.moutassem@aurak.ac.ae}
    \author[1]{Noura 
    Falis}\ead{2023006419@aurak.ac.ae}
    \author[3]{Zainab Islam}\ead{2023006516@aurak.ac.ae}

    \address[1]{Department of Architecture and Civil Engineering, American University of Ras Al Khaimah, Ras Al Khaimah 72603, United Arab Emirates}
    \address[2]{Laboratory of Innovations in Transportation (LiTrans), Toronto Metropolitan University, Canada}
    \address[3]{Department of Computer Science and Engineering, American University of Ras Al Khaimah, Ras Al Khaimah 72603, United Arab Emirates}
    \cortext[cor1]{Corresponding Author.}

\begin{abstract}
This study examines travelers’ preferences for electric flying taxi services in the United Arab Emirates (UAE) under varying travel conditions and service configurations. A stated preference (SP) survey of 213 respondents was conducted to analyze behavior across multiple transport alternatives, including private vehicles, public transport, ground taxis, and both shared and exclusive flying taxi services. The analysis considered key attributes such as travel time and cost, along with contextual factors including travel distance, congestion conditions, day of travel, and trip purpose. In addition, follow-up questions were used to capture the underlying reasons for mode choice and to assess potential modal shifts under changes in travel conditions. The results show that flying taxi services account for 22.6\% of total responses, with higher shares under congested conditions and declining shares as travel distance increases. Clear differences are observed between shared and exclusive services. Shared flying taxis achieve higher modal shares and exhibit greater responsiveness to travel conditions, particularly at moderate distances, during weekdays, and for leisure trips. In contrast, exclusive flying taxis maintain lower modal shares, decline with increasing travel distance, and are more associated with business and weekend travel. The modal shift analysis further indicates that ground taxi users exhibit the highest propensity to switch to shared flying taxi services, particularly under cost increases. These findings highlight the importance of pricing and service design in promoting the adoption of shared flying taxi services as a more sustainable mobility option. In particular, maintaining affordable shared services, ensuring clear price differentiation from exclusive services, and prioritizing deployment in congested corridors and medium-distance travel markets can enhance adoption. The study offers practical insights for the sustainable planning and implementation of urban air mobility systems.
\end{abstract}

\end{frontmatter}

\section{Introduction}
\label{sec:Introduction}
Emerging mobility systems are a growing trend in the transportation sector and are receiving considerable attention from industry and research communities~\citep{narayanan2023shared}. Developments in information and communication technology have contributed to the emergence of various innovative transportation solutions, including electric vehicles, on-demand shared mobility, micromobility, autonomous vehicles, and flying taxis~\citep{alsaleh2023sustainability, xu2025understanding}. As these systems continue to evolve and expand, they are revolutionizing the way people move around cities and communities. Among these, flying taxis are emerging as promising mobility solutions that can support the development of smart and sustainable cities.

Electric flying taxi are generally implemented using electric vertical take-off and landing (eVTOL) aircraft, which form the core technological component of emerging urban air mobility (UAM) systems. eVTOL aircraft use electric propulsion to enable vertical take-off, hovering, and landing capabilities similar to helicopters, while also benefiting from the aerodynamic efficiency of fixed-wing aircraft during cruise flight~\citep{saifudeen2025ongoing}. eVTOL aircraft are also expected to produce lower noise levels and reduced emissions compared with conventional helicopters~\citep{saifudeen2025ongoing, sunitiyoso2025public}.

eVTOL flying taxis are designed to operate as on-demand services within urban environments, providing point-to-point transportation for passengers, cargo, and emergency services between vertiports located at airports, city centers, and other key destinations~\citep{sunitiyoso2025public}. The present study focuses specifically on the passenger transport applications of flying taxis. Most proposed eVTOL air taxi configurations are relatively small aircraft that typically accommodate three to four passengers in addition to a pilot~\cite{akash2021design, doo2021evtol}, although future designs may support fully autonomous operations as automation technologies mature. In terms of performance, many eVTOL concepts are designed for short-to-medium range urban travel, with cruise speeds typically ranging from 120 to 370 km/h~\citep{thipphavong2018urban}, and operational ranges suitable for trips between urban vertiports. Electric flying taxi have the potential to significantly reduce travel times through rapid point-to-point aerial transportation, with door-to-door trip durations estimated to be approximately half or less than those of ground transportation, even when accounting for access and egress travel to vertiports and intermodal transfer times~\citep{antcliff2016silicon, thipphavong2018urban}.

Several countries are planning to introduce electric flying taxis into their urban transport systems. In the United Arab Emirates (UAE), Dubai and Abu Dhabi emirates are planning to launch several fixed-route flying taxi services in 2026~\citep{khaleejtimes2024abudhabi,holtham2024dubai}, followed by Ras Al Khaimah in 2027~\citep{abdulla2024flyingrak}. eVTOL flying taxi systems are increasingly viewed as a promising component of future urban transportation systems aimed at improving accessibility, reducing congestion, and supporting sustainable mobility in rapidly growing cities~\citep{naveen2024unlocking}. Consequently, this technology is expected to offer many benefits that align with the United Nations (UN) Sustainable Development Goals (SDGs), especially SDG 11 (Sustainable Cities) and SDG 13 (Climate Action). 
Despite the growing interest in urban air mobility, empirical evidence on public perceptions and potential demand for flying taxi services remains limited, particularly in regions where the technology is expected to be implemented in the near future. Existing studies, primarily conducted in Europe, North America, and parts of Asia, highlight the importance of travel time savings, cost, and safety perceptions in shaping adoption intentions. However, relatively few studies examine how these factors interact with contextual travel conditions or distinguish between different service configurations, such as shared and exclusive flying taxi services. Moreover, empirical evidence from the Middle East, and specifically the UAE, remains scarce despite the region’s active plans to deploy such systems. Understanding public attitudes and preferences under realistic travel conditions is therefore essential for policymakers and planners to design efficient and sustainable flying taxi services.

Building on this foundation, the current study examines travelers’ preferences for electric flying taxi services in the UAE under varying travel conditions and service configurations, and identifies the key factors influencing their acceptance. The analysis focuses on the effects of travel time, cost, and contextual factors such as travel distance, congestion, day of travel, and trip purpose on mode choice behavior. It distinguishes between shared and exclusive flying taxi services and evaluates their adoption within a broader set of transport alternatives, including private vehicles, public transport, and ground taxis. To achieve this, a stated preference (SP) survey is conducted using hypothetical travel scenarios, complemented by follow-up questions on mode choice reasons and potential modal shifts under changes in travel time and cost.

The remainder of the paper is organized as follows. Section \ref{sec:background} reviews existing studies on travelers’ preferences for electric flying taxi services. Section \ref{sec: Survey & Data} describes the design of the stated preference (SP) survey. Section \ref{sec:results} presents the survey results, including sample characteristics and observed mode choice patterns. Section \ref{sec:dis} discusses the findings, compares them with existing literature, and outlines key recommendations. Finally, Section \ref{sec:conclusions} concludes the paper and highlights directions for future research.

\section{Literature Review}
\label{sec:background}

\subsection{Studying Traveler Preferences for Emerging Mobility Services}

Emerging mobility services such as autonomous vehicles, electric vehicles, shared on-demand mobility services, and urban air mobility are commonly studied using revealed preference (RP) and stated preference (SP) methods. RP surveys are typically applied to mobility services that are already available, allowing researchers to analyze users’ travel behavior and choices under real-world conditions. For example, \cite{alsaleh2023demand} combined observed level-of-service attributes with RP survey data to analyze user behavior toward on-demand transit services. Similarly, \cite{martinez2024electric} used RP data from observed charging sessions to examine electric vehicle users’ charging behavior. In another study, \cite{reck2020much} analyzed travelers’ choices among different mobility-as-a-service (MaaS) subscription plans using RP data.

In contrast, SP surveys are widely used to evaluate new or emerging mobility services prior to their implementation, when RP data are not yet available due to the early stage of technology development \citep{kovzul2025stated}. SP surveys allow researchers to present respondents with hypothetical scenarios describing new transport services and to evaluate their responses under different service attributes, such as travel time, cost, safety, and accessibility. For instance, \cite{tian2021using} conducted a stated choice experiment to examine individuals’ preferences between owning an autonomous vehicle and using shared autonomous car services. Similarly, \cite{pervez2025investigating} employed an SP survey to investigate travelers’ acceptance of autonomous public buses and ride-pooling services. In another study, \cite{adamidis2025urban} used SP data to analyze travelers’ willingness to use and pay for a flying taxi service operating as an airport shuttle.

These studies highlight the importance of SP methods in assessing traveler responses to emerging mobility services and provide useful insights into factors influencing their potential adoption.

\subsection{Traveler Preferences Toward Flying Taxis}

Recent reviews indicate that the flying taxi literature has moved beyond technical feasibility to questions of market potential, public acceptance, safety, noise, security, and integration with existing transport systems. For instance, \cite{tripaldi2025emerging} highlighted that the successful deployment of flying taxi services depends not only on technological readiness but also on the development of supporting infrastructure, suitable regulatory frameworks, and integration with existing transport systems. Another review study by \cite{long2023demand} examined the existing literature on flying taxi demand and identified three main categories of factors influencing potential adoption: trip-related factors (e.g., travel time, cost, and distance), road congestion conditions, and acceptance-related factors such as safety, security, privacy, and noise. Building on this literature, a growing body of research has examined traveler preferences for flying taxis services using survey-based approaches.

Several studies have explored the general acceptance of flying taxis and the factors influencing their adoption. For example, \cite{mostofi2024modelling} examined public attitudes toward flying taxis in Germany and found that expected travel-time savings and congestion reduction positively influence acceptance, whereas concerns related to safety, noise, and increased air traffic negatively affect public perception. The study also reported that travelers with technophilic tendencies were more likely to have a positive attitude toward air taxis. In another recent study, \cite{sunitiyoso2025public} investigated public acceptance of urban air mobility in Jakarta using a SP survey and reported that travel time and service cost are among the most influential attributes affecting travelers’ willingness to use flying taxis. The study also found that perceived that usefulness, trust, and safety significantly influenced individuals’ preference for flying taxi services.

Another group of studies has analyzed traveler preferences using explicit mode choice experiments that compare flying taxis with existing transport options. For example, \cite{fu2019exploring} examined travelers’ preferences for private car, public transportation, autonomous taxi, and autonomous flying taxi in the Munich metropolitan region using SP survey. The results indicated that travel time, cost, and perceived safety influenced travelers’ preferences for both autonomous taxis and flying taxis. The study also reported that younger and older individuals, as well as travelers with higher income levels, were more likely to prefer autonomous flying taxis. In another study, \cite{hwang2023study} developed several operational scenarios in South Korea and compared flying taxis with conventional modes including bus, subway, taxi, and private car. The study considered several service attributes, including access time, waiting time, boarding time, travel cost, autonomous operation, and the existence of other travelers. The findings showed that travel cost and access time significantly affected the likelihood of choosing flying taxis, while the effects of waiting time, boarding time, and autonomous operation varied across the scenarios. More recently, \cite{karimi2024role} examined travelers’ preferences for flying taxis, private cars, and ride hailing services in Tehran, Iran. The analysis considered three trip-related attributes: in-vehicle travel time, travel cost, and access time. The results indicated that travel cost, enjoyment, time-saving benefits, and travelers’ income significantly influenced the preference for flying taxis. 

Other studies have examined specific use cases such as commuting and integrated mobility services. For example, \cite{boddupalli2024mode} conducted a SP survey in the United States to analyze commuters’ choices among private car, public transport, ride hailing, and air taxi services. The study considered several trip-related attributes, including travel cost, in-vehicle travel time, out-of-vehicle travel time, ride guarantee, and transfers, and found that male travelers and frequent ride hailing users were more likely to select air taxi services. \cite{garrow2025market} identified significant heterogeneity in potential demand for flying taxi commuting services across United States cities, with different traveler segments showing varying levels of enthusiasm and concern toward the technology. Furthermore, \cite{zhao2025exploring} conducted a SP survey to examine commuters’ preferences for the integration of flying taxis into Mobility-as-a-Service (MaaS) platforms in Beijing, China. The results indicated that commuters aged 44 and above, higher-income individuals, car owners, regular car users, individuals in managerial roles, and those with helicopter experience were more likely to select multimodal mobility packages that involve flying taxis. The study also found that subscription-based MaaS packages incorporating flying taxis were generally favored over pay-as-you-go alternatives.

Airport-access applications have also been investigated because flying taxis may provide significant travel-time savings for such trips. For example, \cite{adamidis2025urban} examined travelers’ willingness to pay for a flying taxi service operating as an airport shuttle connecting Munich Airport in Germany with key locations in its catchment area. The results indicate that although flying taxis can reduce travel time, many respondents remained satisfied with existing airport access modes, resulting in lower-than-expected willingness to pay. Similarly, \cite{jang2025urban} analyzed airport-access mode choice in South Korea including airport train, airport bus, taxi, private car, and flying taxi. The results found that travelers with higher income levels, frequent taxi users, and those seeking faster airport access are more likely to adopt flying taxi services. \cite{coppola2025urban} conducted a SP survey to examine travelers’ willingness to use flying taxis for airport shuttle services, general air taxi operations, and intercity air travel in the metropolitan area of Milan, Italy. The results indicated that flying taxis were more attractive as airport shuttle services than as general air taxis. Furthermore, the use of flying taxis for intercity trips was influenced by travel distance as well as access and egress times to vertiports.

Overall, existing studies consistently highlight travel time savings, service cost, safety perceptions, and trust as key determinants of flying taxi adoption. However, several gaps remain in the literature.
\begin{itemize}
    \item Many studies focus on specific applications, such as commuting or airport access, and often consider a limited set of competing transportation alternatives.
    \item Relatively few studies simultaneously examine contextual travel factors such as travel distance, congestion conditions, day of travel (weekday versus weekend), and trip purpose (business versus leisure), all of which may significantly influence travelers’ mode choices.
    \item Most previous studies treat flying taxis as a single service type and rarely distinguish between exclusive and shared flying taxi services within the same choice framework.
    \item Existing empirical evidence is largely concentrated in Europe, the United States, and parts of South and East Asia, with limited research conducted in the Middle East. To the best of our knowledge, no prior study has examined traveler preferences for flying taxi services in the United Arab Emirates (UAE), despite the region’s active plans to deploy such systems in the near future.
\end{itemize}

To address these gaps, the present study makes several contributions to the literature. First, it examines traveler preferences under multiple travel conditions, including travel distance, congestion level, weekday versus weekend travel, and trip purpose, while incorporating key service attributes such as travel time and cost. Second, it evaluates flying taxi adoption within a broader mode choice framework by comparing conventional transport modes, including private vehicles, public transport, and ground taxis, with both exclusive and shared flying taxi services. In particular, the distinction between shared and exclusive services provides insights into how pricing and service configuration influence user choices. Third, the study incorporates follow-up questions to capture the underlying reasons for mode choice and to assess potential modal shifts under changes in travel time and cost, offering a more comprehensive understanding of travelers’ decision-making behavior. Finally, it provides the first empirical investigation, to the best of our knowledge, of traveler preferences for flying taxi services in the UAE context.

\section{Survey Design and Data Analysis}
\label{sec: Survey & Data}

\begin{figure}[h]
\centering
\includegraphics[width=0.9\textwidth]{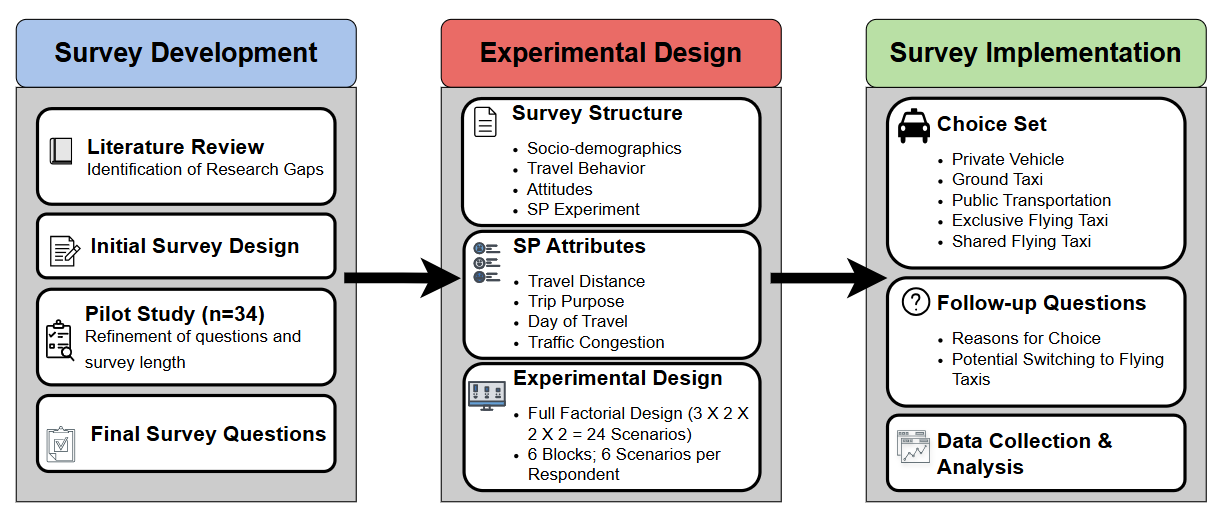}
\caption{Framework of the SP survey design process}
\label{fig:framework}
\end{figure}

This section describes the design of the SP survey used to examine travelers’ preferences toward flying taxi services in the UAE and methods used to analyze the results. The survey was developed to collect information on respondents’ socio-demographic characteristics, current travel behavior, attitudes toward technology and travel costs, and their mode choices under hypothetical travel scenarios. The survey was administered online using the SurveyMonkey platform and targeted individuals currently residing in the UAE. A conceptual overview of the study framework is presented in Figure~\ref{fig:framework}.

\subsection{Survey Structure}
The Survey consisted of four main sections. Each section was designed to serve a distinct purpose in supporting the descriptive and exploratory analysis of travelers’ preferences toward flying taxi services.

The first section collected respondents’ socio-demographic characteristics, including gender, marital status, age group, monthly income, education level, employment status, household size, vehicle ownership, and emirate of residence. These variables were included to describe the sample and to explore whether preferences toward flying taxi services vary across different population groups.

The second section focused on respondents’ current travel behavior. Participants were asked about their primary mode of travel during weekdays and weekends, the average duration of their one-way work trip, monthly transportation expenditure, frequency of international air travel, and exposure to traffic congestion, measured as the additional time added to daily trips due to congestion. Respondents were also asked about the frequency of intercity travel between Emirates, the modes typically used for these trips, and their trip purposes. This section was included to provide context on current mobility patterns and to assess how existing travel behavior may be related to preferences for flying taxi services.

The third section included attitudinal statements designed to measure respondents’ trust in technology and sensitivity to travel costs. Trust in technology was assessed using statements related to the adoption of new technologies and the use of smart technological features in everyday life. Cost sensitivity was measured using statements reflecting the importance of travel price in decision-making and willingness to pay for convenience or time savings. Responses were recorded using a five-point Likert scale ranging from strongly disagree to strongly agree.  The attitudinal statements used in the survey are summarized in Table~\ref{tab:attitudes_statements}.

\begin{table}[ht]
\caption{Statements used to assess attitudes toward technology and cost sensitivity}
\centering
\begin{tabularx}{\textwidth}{@{}lX@{}}
\hline
\textbf{Attitudes} & \textbf{Statements} \\
\hline
\multirow{4}{*}{Trust in Technology} 
& I enjoy trying new technologies before they become widely adopted. \\
& I usually wait until others have tested a new technology before I try it myself. \\
& I have doubts about whether new technologies actually improve everyday life. \\
& I regularly use smart features in my car, phone, or home. \\
\hline
\multirow{4}{*}{Cost Sensitivity} 
& Price is a major factor in my travel decisions. \\
& I am willing to pay more for convenience or time savings. \\
& I often compare prices before deciding how to travel. \\
& I avoid using transport options that I consider too expensive. \\
\hline
\end{tabularx}
\label{tab:attitudes_statements}
\end{table}

The fourth section contained the SP experiment. In this section, respondents were presented with hypothetical travel scenarios and asked to choose their preferred transportation mode from a predefined set of alternatives. This section formed the core of the survey and was designed to evaluate travelers’ preferences toward flying taxi services under different travel conditions.

\subsection{Stated Preference Experiment}
The SP experiment was designed to examine how travelers preferences for conventional transport modes and flying taxi services under different trip contexts. In each choice task, respondents selected one alternative from a choice set containing five transportation options: private vehicle, ground taxi, public transportation, exclusive flying taxi, and shared flying taxi. The inclusion of both conventional modes and two flying taxi service configurations allowed the survey to assess not only the attractiveness of flying taxis relative to existing options, but also the potential differences in respondents’ preferences for exclusive versus shared flying taxi services.

The selection of attributes and their corresponding levels was guided by previous studies investigating traveler preferences for flying taxi services. Prior research has identified travel time and travel cost as key factors influencing travelers’ willingness to adopt flying taxi services \citep{fu2019exploring, hwang2023study, karimi2024role}. In addition, contextual travel characteristics such as traffic congestion \citep{long2023demand} and travel distance \citep{long2023demand, coppola2025urban} have been shown to influence travelers' preference for air taxis. Furthermore, trip purpose has been found to affect the adoption of emerging mobility services more broadly, including autonomous vehicles \citep{ashkrof2019impact}, ride hailing services \citep{hossain2021inferring}, and on-demand transit services \citep{alsaleh2023demand}. Therefore, the attributes included in this study were selected to represent realistic travel conditions in the UAE while capturing key factors expected to influence travelers’ mode choice behavior.

Four contextual attributes were used to define the hypothetical travel scenarios: travel distance, trip purpose, day of travel, and traffic congestion. These attributes and their levels are presented in Table~\ref{tab:attributes_levels}. In addition, travel time and travel cost were specified for each transportation alternative within every scenario. Unlike the contextual attributes used to construct the factorial design, travel time and travel cost were treated as alternative-specific attributes whose values varied across modes and scenarios. These values were determined based on the scenario conditions, particularly travel distance, traffic congestion, and the day of travel, to ensure that the presented travel options reflected realistic travel conditions.

\begin{table}[ht]
\centering
\caption{Attributes and levels used to construct the SP scenarios}
\label{tab:attributes_levels}
\begin{tabular}{p{4cm} p{9cm}}
\hline
Attribute & Levels \\
\hline
Travel distance & 22 km, 76 km, 140 km \\
Trip purpose & Business, Leisure \\
Day of travel & Weekday, Weekend \\
Traffic congestion & Congested, Uncongested \\
\hline
\end{tabular}
\end{table}

The selected travel distances correspond to representative trips within the UAE. The short-distance trip represented intra-city travel within Dubai (Downtown Dubai to Dubai Marina), the medium-distance trip represented inter-emirate travel between Dubai and Sharjah (Downtown Dubai to Al Dhaid), and the long-distance trip represented travel between Dubai and Abu Dhabi. Trip purpose was defined as either business or leisure, while day of travel was distinguished between weekday and weekend conditions. Traffic congestion was represented using two levels: congested and uncongested. A full factorial design was used to generate the SP scenarios. Combining the four contextual attributes resulted in a total of 24 possible scenarios $(3 \times 2 \times 2 \times 2 = 24)$.

For each scenario, travel time and travel cost values were specified for all five transportation alternatives. Travel times for conventional modes (private vehicle, ground taxi, and public transportation) were estimated using the Google Maps platform~\citep{google_maps_uae} under the relevant traffic and day of week conditions. Taxi costs were estimated using the fare structure of the Hala Taxi platform in Dubai (the main taxi e-hailing service in Dubai and Ras Al Khaimah, operated through the Careem platform)~\citep{hala_taxi_fare}. Private vehicle costs were estimated using an online vehicle operating cost calculator~\citep{rome2rio_trip_planner} and adjusted according to travel distance and travel time, while public transportation costs were based on applicable local fare structures~\citep{rta_intercity_bus_booking, rta_nol_fares}.

\begin{figure}[H]
    \centering
 \includegraphics[width=0.75\textwidth]{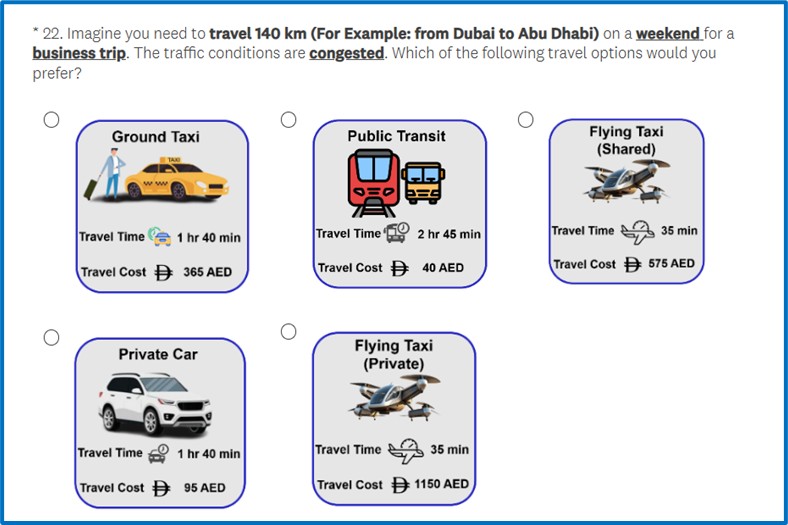}
 \caption{Example of SP choice scenarios presented to respondents.}
 \label{Fig2}
\end{figure}

For flying taxi services, travel time estimates were derived from publicly available information on planned flying taxi operations in the UAE~\citep{khaleejtimes_flying_taxi_abu_dhabi_dubai, tiketi_air_taxi_dubai}. The cost of exclusive flying taxi services was estimated based on price ranges reported in regional announcements and news sources related to future flying taxi operations~\citep{tiketi_air_taxi_dubai}. The cost of shared flying taxi services was assumed to be 50\% of the exclusive flying taxi fare, while travel time was assumed to remain unchanged. This assumption was adopted to represent a lower-cost shared flying taxi option. Figure~\ref{Fig2} presents an example of a SP choice task as shown to respondents.

\subsection{Experimental Design and Scenario Construction}
The 24 scenarios were divided into six blocks using a blocking strategy to reduce respondent burden and maintain engagement. Two blocks were defined for each travel distance, with one block representing business travel scenarios and the other representing leisure travel scenarios. Within each block, the day of travel and congestion conditions varied across four scenarios. Each respondent was randomly assigned six choice tasks, one from each block. As a result, every participant evaluated two scenarios for each travel distance: one business trip scenario and one leisure trip scenario. The complete list of the 24 scenarios, including the corresponding travel time and cost values for all alternatives is provided in Table~\ref{tab:scenarios}.

Following each choice task, respondents were asked two follow-up questions. First, they were asked to indicate the main reason for selecting their preferred mode from a predefined set of options. Second, they were asked whether they would consider switching to a flying taxi service if relevant conditions changed, such as travel time or travel distance. The exact wording and response options of these follow-up questions depended on the transportation mode selected by the respondent. 

An attention-check question was embedded within the SP section of the survey to ensure data quality and respondent attentiveness. Specifically, after completing three choice scenarios, respondents were presented with a simple instruction-based question asking them to select a specific response (e.g., \textit{“please select ‘4’ for this question”}). This type of validation question is commonly used in survey-based research to screen for inattentive or random answering.

\subsection{Pilot Study}

Prior to the full deployment of the survey, a pilot study was conducted to assess the clarity, comprehensibility, and overall length of the survey. The pilot survey resulted in 34 completed responses. Feedback from the pilot respondents was used to refine the survey in two main ways. First, several questions were revised to improve clarity, particularly by replacing technical or unfamiliar terms with simpler wording. Second, a number of non-essential questions were removed in order to reduce respondent burden and maintain a manageable survey duration.

These revisions shortened the final survey to approximately 10 minutes, which was considered appropriate for minimizing fatigue while still collecting the information necessary to support the study objectives. The pilot responses also indicated that the SP scenarios and the overall structure of the survey were generally understandable to respondents.

\subsection{Data Analysis}

The analysis conducted in this study focuses on descriptive statistics and exploratory analysis of the SP responses to provide initial insights into travelers’ preferences toward flying taxi services. The analysis is primarily based on frequency distributions, percentages, and cross-tabulations of modal choices across different SP scenarios and travel attributes, including travel distance, trip purpose, day of travel, and traffic congestion levels. In addition, aggregate and disaggregate modal shares are examined to identify patterns in respondents’ preferences, with particular focus given to the variation between exclusive and shared flying taxi services. 

It is important to note that the present study does not estimate discrete choice models. The development of random utility maximization (RUM)-based models, such as mixed logit models, is beyond the scope of this paper and and will be developed in future work to further quantify preferences. In addition to descriptive analysis, Pearson's chi-square tests of independence~\citep{pearson1900criterion} are conducted to assess whether the observed differences in mode choice distributions across key travel attributes (travel distance, traffic congestion, trip purpose, and day of travel) are statistically significant. These tests evaluate the null hypothesis that mode choice is independent of the attribute under consideration, providing formal statistical support for the descriptive patterns reported in the results. 

\section{Results}
\label{sec:results}
The survey was administered online using the SurveyMonkey platform over a period of approximately five weeks, from November 25, 2025, to January 1, 2026. The target population included individuals aged 18 years and above residing in the UAE. The survey was open to the public to ensure a diverse sample in terms of socio-demographic characteristics and travel behavior.

The survey protocol was reviewed and approved by the Institutional Review Board (IRB) of the authors’ institution. Participation was voluntary, and informed consent was obtained from all respondents prior to data collection. Respondents were informed of the purpose of the study and their right to withdraw at any time. To ensure privacy and confidentiality, no personally identifiable information was collected, and all responses were anonymized. The data were used solely for research purposes, and no disaggregated information that could identify individual participants is reported.

A total of 273 responses were received. Several screening criteria were applied to ensure the reliability and validity of the dataset. Responses were excluded if they were incomplete, if the respondent was younger than 18 years, if the respondent resided outside the UAE, or if the data quality check question was answered incorrectly. Following this data cleaning process, 213 valid responses were retained for analysis.

This section presents the empirical findings of the study, including the characteristics of the sample, observed travel behavior patterns, and the results of the SP experiment.

\subsection{Sample Characteristics and Travel Behavior Patterns}
Understanding the socio-demographic characteristics of respondents is important for interpreting travel preferences and the potential adoption of emerging mobility services. Table~\ref{tab:socio_demo} summarizes the key characteristics of the survey sample. The sample comprises a slightly higher percentage of male respondents (57.3\%) compared to female respondents (42.7\%), with the vast majority being either married or single (98.1\%). The age distribution is largely concentrated within the economically active population, with approximately 80\% of respondents aged between 18 and 44 years. This demographic profile is particularly relevant in the context of emerging mobility systems, as younger and middle-aged individuals have been shown to be more likely to adopt new transport technologies, including autonomous vehicles~\citep{clayton2020autonomous}, on-demand transit~\citep{alsaleh2023demand}, and ride hailing services~\citep{young2019and}.

\footnotesize
\begin{longtable}{>{\raggedright\arraybackslash}p{3.4cm}
                  >{\raggedright\arraybackslash}p{6.0cm}
                  c c}
\caption{Socio-demographic characteristics of survey respondents}
\label{tab:socio_demo} \\

\toprule
\textbf{Characteristic} & \textbf{Category} & \textbf{Count (n)} & \textbf{Percentage (\%)} \\
\midrule
\endfirsthead

\caption[]{Socio-demographic characteristics of survey respondents (continued)} \\
\toprule
\textbf{Characteristic} & \textbf{Category} & \textbf{Count (n)} & \textbf{Percentage (\%)} \\
\midrule
\endhead

\midrule
\multicolumn{4}{r}{\textit{Continued on next page}} \\
\endfoot

\bottomrule
\endlastfoot

Gender & Male & 122 & 57.3 \\
 & Female & 91 & 42.7 \\

\addlinespace
Marital Status & Single & 96 & 45.1 \\
 & Married & 113 & 53.0 \\
 & Divorced & 4 & 1.9 \\
 & Widowed & 0 & 0.0 \\

\addlinespace
Age Group & 18--24 & 68 & 31.9 \\
 & 25--34 & 47 & 22.0 \\
 & 35--44 & 57 & 26.8 \\
 & 45--54 & 33 & 15.5 \\
 & 55--64 & 7 & 3.3 \\
 & 65+ & 1 & 0.5 \\

\addlinespace
Education & Less than high school & 0 & 0.0 \\
 & High school diploma or equivalent & 50 & 23.5 \\
 & College or associate degree & 8 & 3.7 \\
 & Bachelor's degree & 69 & 32.4 \\
 & Master's degree & 53 & 24.9 \\
 & Doctorate (PhD) & 33 & 15.5 \\

\addlinespace
Employment Status & Employed full-time & 117 & 54.9 \\
 & Employed part-time & 16 & 7.5 \\
 & Self-employed & 17 & 8.0 \\
 & Unemployed & 8 & 3.7 \\
 & Student & 53 & 24.9 \\
 & Retired & 2 & 1.0 \\

\addlinespace
Monthly Income (AED) & Less than 5,000 & 36 & 16.9 \\
 & 5,000--9,999 & 28 & 13.2 \\
 & 10,000--14,999 & 29 & 13.6 \\
 & 15,000--19,999 & 38 & 17.8 \\
 & 20,000--29,999 & 50 & 23.5 \\
 & 30,000--49,999 & 25 & 11.7 \\
 & 50,000 or more & 7 & 3.3 \\

\addlinespace
Vehicle Ownership & Yes & 167 & 78.4 \\
 & No & 46 & 21.6 \\

\addlinespace
Household Size & 1 Person & 15 & 7.0 \\
 & 2 Persons & 14 & 6.6 \\
 & 3 Persons & 41 & 19.3 \\
 & 4 Persons & 45 & 21.1 \\
 & 5 Persons & 59 & 27.7 \\
 & 6 Persons & 25 & 11.7 \\
 & 7 or more & 14 & 6.6 \\

\addlinespace
Residency & Abu Dhabi & 26 & 12.2 \\
 & Dubai & 65 & 30.5 \\
 & Sharjah & 22 & 10.3 \\
 & Ajman & 24 & 11.3 \\
 & Umm Al Quwain & 6 & 2.8 \\
 & Ras Al Khaimah & 67 & 31.5 \\
 & Fujairah & 3 & 1.4 \\

\end{longtable}

The sample also exhibits a high level of education, with more than 70\% of respondents holding at least a bachelor’s degree. The employment distribution reflects a predominantly active population, with 54.9\% employed full-time and 24.9\% identified as students. This is particularly relevant to the present study, as employment status is closely associated with travel frequency, value of time, and the propensity to adopt emerging mobility services.

Income distribution reflects a moderate-to-high earning population, with the largest group (23.5\%) earning between 20,000 and 29,999 AED per month (approximately 5,444–-8,166 USD). Previous research has highlighted income as a key determinant in the adoption of premium mobility services, indicating that higher-income individuals are more likely to consider flying taxis due to their ability to afford higher travel costs~\citep{fu2019exploring, karimi2024role}. 

Car ownership is notably high, with 78.4\% of respondents reporting ownership or leasing of a private vehicle. This pattern suggests a strong reliance on private vehicles among the surveyed population and is consistent with the travel behavior patterns observed in Table~\ref{tab:travel_behaviour}. The household size distribution indicates that 67.1\% of respondents reside in households comprising four or more persons. Larger household sizes may influence travel behavior, particularly in terms of cost sensitivity and the potential attractiveness of shared mobility options.

In terms of geographic distribution, respondents are drawn from multiple emirates, with the largest shares residing in Ras Al Khaimah (31.5\%) and Dubai (30.5\%), followed by Abu Dhabi, Ajman, and Sharjah. This distribution captures a mix of urban and semi-urban contexts, which is important because travel patterns, congestion levels, and accessibility to transport infrastructure vary across emirates. Such variation provides a broader perspective on the potential adoption of flying taxi services across different spatial contexts within the UAE.

Table~\ref{tab:travel_behaviour} presents the key travel behavior characteristics of the survey respondents. The results indicate that private vehicles represent the dominant mode of travel for both weekdays (73.7\%) and weekends (75.6\%), while public transport and ride hailing services account for relatively smaller shares (18.8\% and 18.4\% on weekdays and weekends, respectively). Active modes, such as walking and micromobility, are used by a limited proportion of respondents (4.2\% and 2.2\% on weekdays and weekends, respectively). In terms of commuting patterns, approximately half of the respondents report one-way travel times of less than 30 minutes, while a notable share (36.1\%) experiences longer commute durations exceeding 30 minutes, including 5.3\% who report commuting times of more than 90 minutes. 

\begin{table}[!htbp]
\centering
\caption{Travel behavior characteristics of survey respondents}
\label{tab:travel_behaviour}
\footnotesize
\begin{threeparttable}

\begin{tabular}{>{\raggedright\arraybackslash}p{3.4cm}
                >{\raggedright\arraybackslash}p{6.0cm}
                c c}
\toprule
\textbf{Characteristic} & \textbf{Category} & \textbf{Count (n)} & \textbf{Percentage (\%)} \\
\midrule

\multirow[t]{6}{=}{\makecell[tl]{Primary Mode of\\Travel During Weekdays}}
 & Private car & 157 & 73.7 \\
 & Carpool/shared ride with friends or colleagues & 7 & 3.3 \\
 & Taxi or ride-hailing apps (e.g., Careem) & 18 & 8.5 \\
 & Public transport (bus/metro) & 22 & 10.3 \\
 & Bicycle/e-scooter & 0 & 0.0 \\
 & Walking & 9 & 4.2 \\

\addlinespace
\multirow[t]{6}{=}{\makecell[tl]{Primary Mode of\\Travel During Weekends}}
 & Private car & 161 & 75.6 \\
 & Carpool/shared ride with friends or colleagues & 8 & 3.8 \\
 & Taxi or ride-hailing apps (e.g., Careem) & 21 & 9.9 \\
 & Public transport (bus/metro/train) & 18 & 8.5 \\
 & Bicycle/e-scooter & 2 & 0.9 \\
 & Walking & 3 & 1.3 \\

\addlinespace
\multirow[t]{6}{=}{\makecell[tl]{Average One-Way\\Commuting Time}}
 & Not applicable (unemployed) & 16 & 7.5 \\
 & Less than 15 minutes & 61 & 28.6 \\
 & 15--29 minutes & 48 & 22.5 \\
 & 30--59 minutes & 42 & 19.7 \\
 & 60--89 minutes & 35 & 16.4 \\
 & More than 90 minutes & 11 & 5.3 \\

\addlinespace
\multirow[t]{6}{=}{\makecell[tl]{Monthly Transportation\\Expenditure (AED)}}
 & Less than 200 & 39 & 18.3 \\
 & 200--499 & 48 & 22.5 \\
 & 500--999 & 51 & 23.9 \\
 & 1,000--1,999 & 52 & 24.4 \\
 & 2,000--3,999 & 21 & 9.9 \\
 & 4,000 or more & 2 & 1.0 \\

\addlinespace
\multirow[t]{6}{=}{\makecell[tl]{Annual International\\Air Travel Frequency}}
 & Never & 30 & 14.1 \\
 & 1--2 times & 102 & 47.9 \\
 & 3--4 times & 46 & 21.6 \\
 & 5--7 times & 24 & 11.3 \\
 & 8--10 times & 5 & 2.3 \\
 & More than 10 times & 6 & 2.8 \\

\addlinespace
\multirow[t]{6}{=}{\makecell[tl]{Inter-Emirate Travel\\Frequency}}
 & Daily & 31 & 14.6 \\
 & Several times a week & 18 & 8.5 \\
 & Weekly & 44 & 20.7 \\
 & Monthly & 56 & 26.3 \\
 & Occasionally & 56 & 26.3 \\
 & Never & 8 & 3.6 \\

\addlinespace
\multirow[t]{5}{=}{\makecell[tl]{Primary Intercity\\Travel Mode}}
 & Private car & 139 & 65.3 \\
 & Intercity bus & 32 & 15.0 \\
 & Taxi or ride-hailing & 30 & 14.1 \\
 & Carpool or shared ride & 10 & 4.7 \\
 & Other & 2 & 0.9 \\

\addlinespace
\multirow[t]{5}{=}{\makecell[tl]{Primary Intercity\\Travel Purpose}}
 & Work or business & 77 & 36.2 \\
 & Family or personal visits & 68 & 31.9 \\
 & Tourism or leisure & 41 & 19.3 \\
 & Medical or official appointments & 11 & 5.2 \\
 & Other & 16 & 7.4 \\

\addlinespace
\multirow[t]{6}{=}{\makecell[tl]{Daily Congestion\\Delay (Minutes)}}
 & Less than 10 & 43 & 20.2 \\
 & 10--19 & 46 & 21.6 \\
 & 20--29 & 52 & 24.4 \\
 & 30--44 & 40 & 18.8 \\
 & 45 or more & 21 & 9.9 \\
 & Not sure & 11 & 5.1 \\

\bottomrule
\end{tabular}

\end{threeparttable}
\end{table}

Monthly transportation expenditure is distributed across multiple categories, with most respondents reporting spending between 200 and 1,999 AED. The findings further show that a majority of respondents travel internationally at least once per year, with nearly half reporting one to two trips annually. Inter-emirate travel appears to be relatively frequent, with 43.8\% of respondents traveling across emirates on a weekly basis. For intercity trips, private vehicles represent the most commonly used mode (65.3\%), followed by intercity buses (15\%) and ride-hailing services (14.1\%). Work or business trips are reported to be the most common purpose of intercity travel (36.2\%), followed by personal (31.9\%) and leisure-related trips (19.3\%). Finally, 20.2\% of respondents report minimal daily congestion delays of less than 10 minutes, while 53.1\% experience delays exceeding 20 minutes, including 9.9\% who report delays of more than 45 minutes.

\subsection{Modal Share Variation Across Travel Attributes}

The SP experiments were designed to examine how respondents adjust their mode choices under varying travel conditions. A detailed breakdown of modal shares across all 24 scenarios is provided in Table~\ref{tab:mode_choice_distribution}.

\begin{figure}[!htbp]
    \centering
    \includegraphics[width=0.85\textwidth]{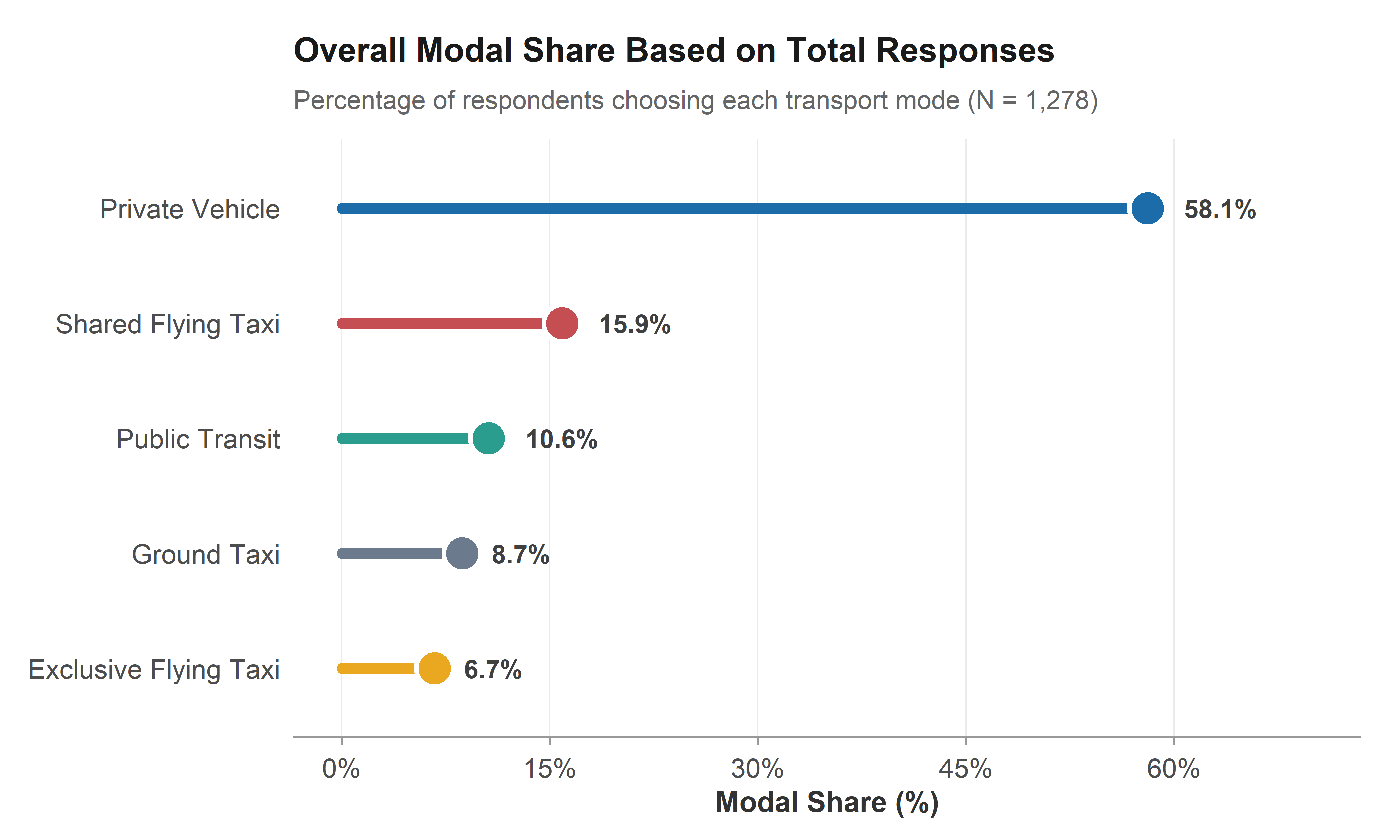}
    \caption{Modal share based on total responses across all scenarios}
    \label{Fig3}
\end{figure}

The overall modal distribution, based on 1,278 observations obtained from 213 respondents across six SP choice tasks, is presented in Figure~\ref{Fig3}. Private vehicles represent the predominant mode, accounting for 58.1\% of total choices. Public transport and ground taxi services account for comparatively smaller shares of 10.6\% and 8.7\%, respectively. When flying taxi alternatives are considered jointly, the combined share of exclusive and shared services amounts to 22.6\% of total responses. Within this category, shared flying taxis account for a notably larger proportion (15.9\%), whereas exclusive flying taxis represent 6.7\%, which corresponds to the lowest modal share among the available alternatives. These aggregate results establish a baseline representation of modal preferences across the full set of scenarios.

\begin{figure}[!htbp]
    \centering
    \includegraphics[width=0.85\textwidth]{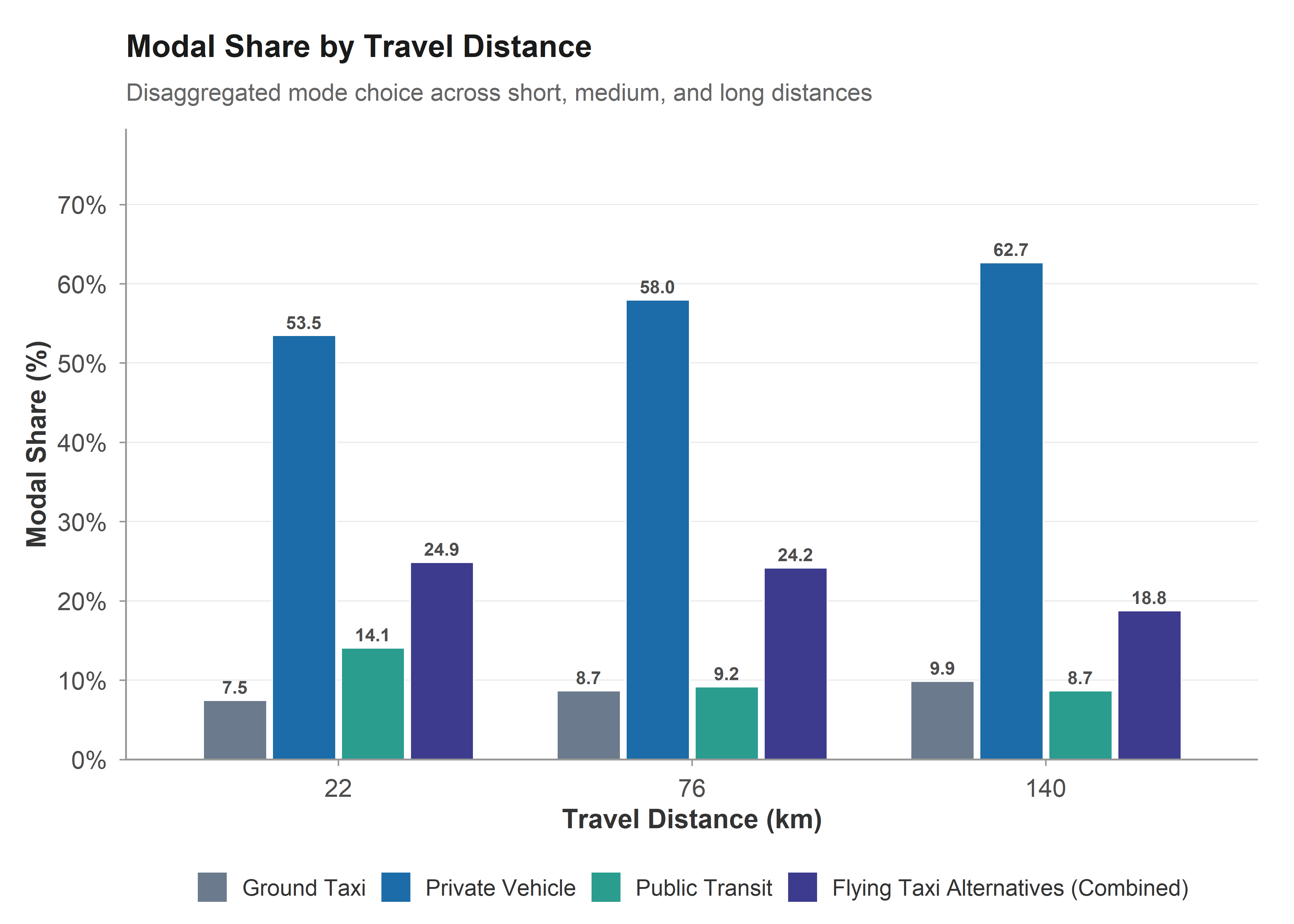}
    \caption{Modal share distribution by travel distance, illustrating variations in mode choice across short (22 km), medium (76 km), and long (140 km) trips.}
    \label{Fig4}
\end{figure}

\begin{figure}[!htbp]
    \centering
    \includegraphics[width=0.85\textwidth]{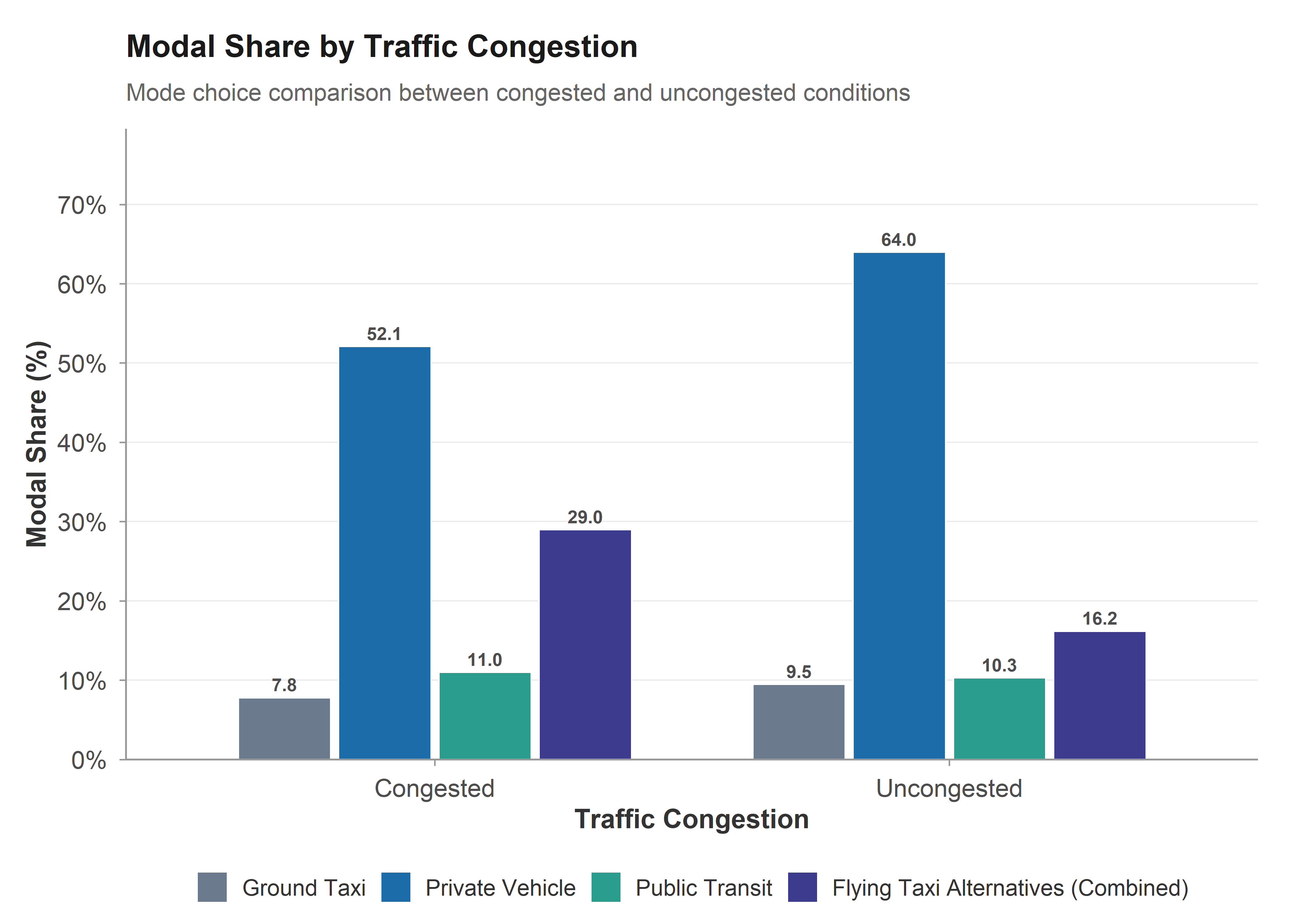}
    \caption{Modal share distribution by traffic congestion}
    \label{Fig5}
\end{figure}

\begin{figure}[!htbp]
    \centering
    \includegraphics[width=0.85\textwidth]{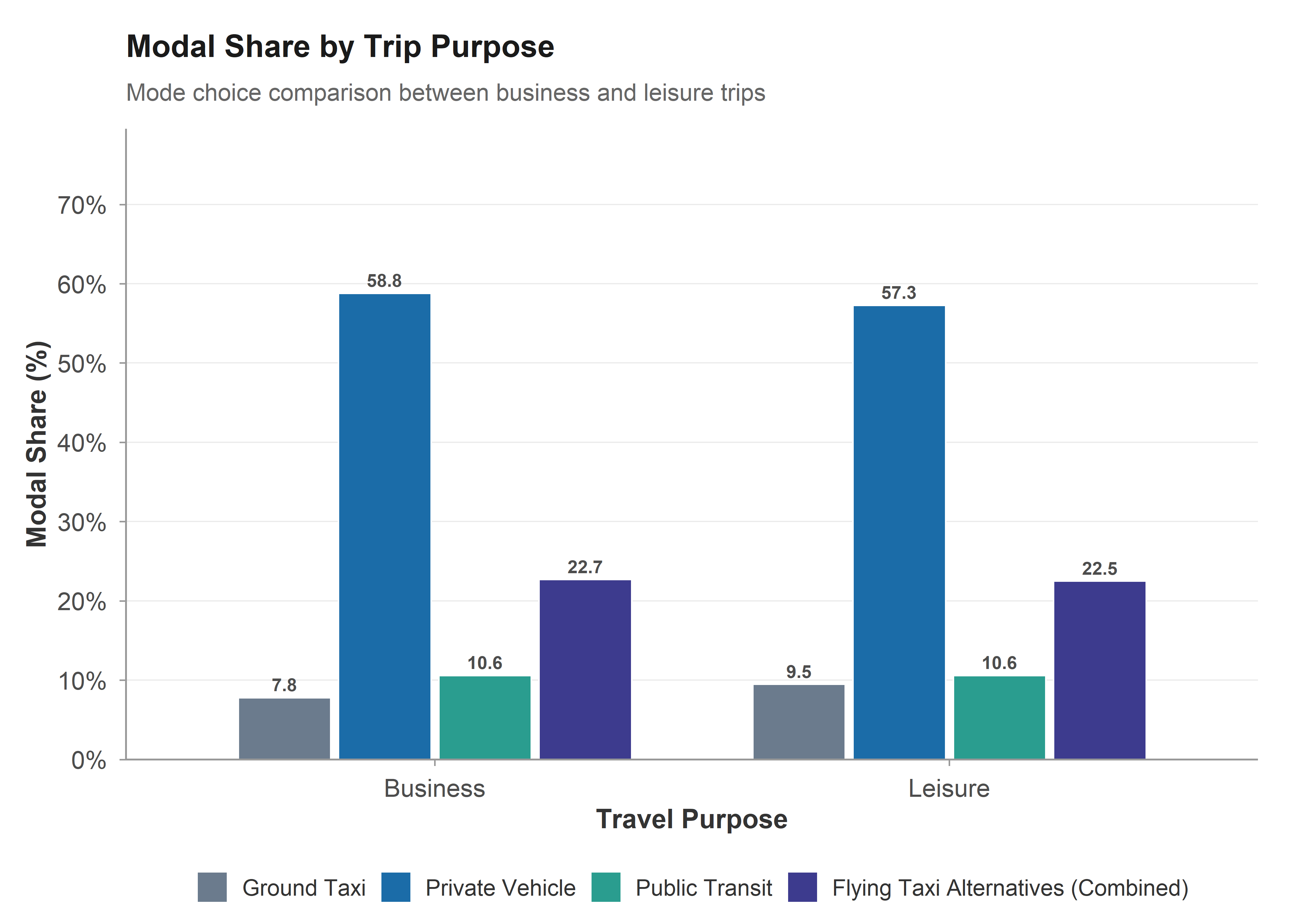}
    \caption{Modal share distribution by trip purpose}
    \label{Fig6}
\end{figure}

\begin{figure}[!htbp]
    \centering
    \includegraphics[width=0.85\textwidth]{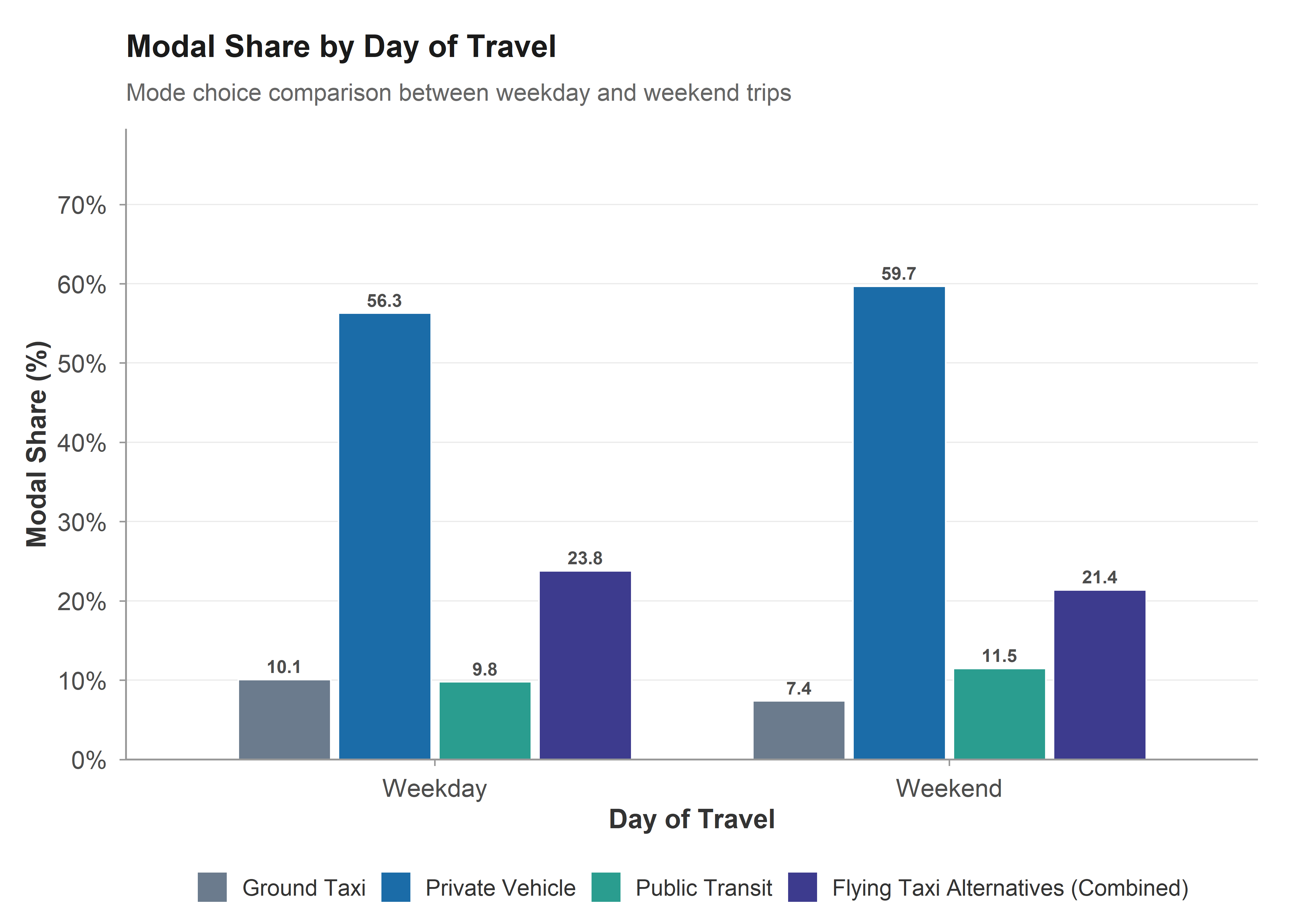}
    \caption{Modal share distribution by day of travel}
    \label{Fig7}
\end{figure}

Figures~\ref{Fig4}--\ref{Fig7} present the variation in modal share across the four key attributes considered in the SP experiment, namely travel distance, trip purpose, day of travel, and traffic congestion. The figures report the modal distribution for each transport alternative, including the combined share of flying taxi services (exclusive and shared).

Figure~\ref{Fig4} illustrates that travel distance is associated with a clear variation in modal preferences. The combined share of flying taxi alternatives remains relatively stable for short and moderate distances, accounting for 24.9\% at 22 km and 24.2\% at 76 km. However, this share decreases notably to 18.8\% for the longer distance of 140 km. This pattern indicates that flying taxi services are more attractive for short-to-moderate intercity trips, while their relative competitiveness declines for longer-distance travel. This decline is accompanied by a steady increase in private vehicle usage, which rises from 53.5\% to 58.0\% and 62.7\% across the three distances. Public transport usage decreases from 14.1\% at shorter distances to 8.7\% at longer distances, while ground taxi usage increases slightly from 7.5\% to 9.9\%. 

These trends can be explained by the combined effects of travel cost and travel time. As travel distance increases, the cost associated with flying taxi alternatives becomes more significant, reducing their relative attractiveness despite their shorter travel times. At the same time, the longer travel times associated with public transport further reduce its competitiveness. As a result, respondents tend to shift toward modes that provide a more favorable balance between cost and travel time, particularly private vehicles and ground taxis, which are generally less expensive than flying taxi alternatives while offering shorter travel times than public transport.

Traffic congestion is associated with the most pronounced variation in modal distribution, as shown in Figure~\ref{Fig5}. The overall flying taxi share increases substantially from 16.2\% under uncongested conditions to 29.0\% under congested conditions. This increase is accompanied by a notable reduction in private vehicle usage from 64.0\% to 52.1\%. Public transport usage shows a slight increase from 10.3\% to 11.0\%, while ground taxi usage decreases from 9.5\% to 7.8\%. These results highlight the strong influence of traffic congestion on flying taxi adoption, with higher uptake observed when road-based travel becomes more constrained. 

This pattern can be explained by the differential impact of congestion on travel time across modes. Under congested conditions, the travel time of conventional modes increases substantially, which reduces their relative attractiveness despite their lower cost. In contrast, the travel time advantage offered by flying taxi alternatives becomes more pronounced, which enhances their competitiveness and encourages modal shifts from conventional modes. At the same time, for cost-sensitive travelers who may not be able to afford flying taxi alternatives, public transport becomes a relatively more attractive option, as its travel time becomes more comparable to other surface modes while maintaining lower costs. This explains the modest increase observed for the public transport alternative.

In contrast, trip purpose exhibits limited influence on overall modal distribution (Figure~\ref{Fig6}). The overall flying taxi share remains nearly constant at 22.7\% for business trips and 22.5\% for leisure trips. Private vehicle usage accounts for 58.8\% of choices for business trips and 57.3\% for leisure trips, while public transport remains unchanged at 10.6\% for both purposes. Ground taxi usage increases slightly from 7.8\% for business trips to 9.5\% for leisure trips. These findings suggest that trip purpose has a minimal effect on the overall adoption of flying taxi services and primarily results in minor redistribution across other modes. A similarly modest effect is observed for the day of travel (Figure~\ref{Fig7}). The overall flying taxi share decreases slightly from 23.8\% on weekdays to 21.4\% on weekends. Private vehicle usage increases from 56.3\% to 59.7\%, while public transport usage increases from 9.8\% to 11.5\%. In contrast, ground taxi usage decreases from 10.1\% to 7.4\%. Overall, these variations indicate that the day of travel has a limited influence on aggregate modal preferences. 

The patterns observed for day of travel and trip purpose attributes can be explained by the observation that, unlike travel distance and traffic conditions, which directly affect both travel time and cost, trip purpose and day of travel do not substantially alter these attributes within the experimental design. Consequently, mode choice behavior remains relatively stable across different trip purposes and days of travel, with only marginal redistribution among modes.

To provide statistical support for the observed differences in modal share across travel attributes, Pearson's chi-square tests of independence were conducted. These tests assess whether the distribution of mode choices is independent of each contextual attribute, using the aggregate counts across all scenarios within each attribute level. The results are summarized in Table~\ref{tab:chi_square}.
 
\begin{table}[ht]
\centering
\caption{Chi-square tests of independence for mode choice distribution across scenario attributes}
\label{tab:chi_square}
\begin{tabular}{lccc}
\toprule
\textbf{Attribute} & \textbf{$\chi^2$} & \textbf{df} & \textbf{$p$-value} \\
\midrule
\multicolumn{4}{l}{\textit{Five-mode distribution}} \\
\addlinespace
Traffic congestion & 35.294 & 4 & $<$0.001*** \\
Travel distance & 27.219 & 8 & $<$0.001*** \\
Trip purpose & 7.654 & 4 & 0.105 \\
Day of travel & 8.652 & 4 & 0.070 \\
\addlinespace
\multicolumn{4}{l}{\textit{Four-category distribution (flying taxi alternatives combined)}} \\
\addlinespace
Traffic congestion & 32.097 & 3 & $<$0.001*** \\
Travel distance & 15.789 & 6 & 0.015* \\
Trip purpose & 1.228 & 3 & 0.746 \\
Day of travel & 5.084 & 3 & 0.166 \\
\bottomrule
\multicolumn{4}{l}{\footnotesize *$p < 0.05$; **$p < 0.01$; ***$p < 0.001$}
\end{tabular}
\end{table}
 
The chi-square results confirm that traffic congestion and travel distance produce statistically significant differences in modal distribution ($p < 0.001$ for both attributes under the five-mode specification). In contrast, trip purpose ($p = 0.105$) and day of travel ($p = 0.070$) do not yield statistically significant differences at the 5\% level, consistent with the limited variation observed in Figures~\ref{Fig6} and~\ref{Fig7}. When flying taxi alternatives are combined into a single category, the same pattern holds: traffic congestion ($p < 0.001$) and travel distance ($p = 0.015$) remain significant, while trip purpose and day of travel remain non-significant. These results statistically validate the descriptive patterns reported above and confirm that congestion and distance are the primary contextual drivers of modal variation in this study.

\subsection{Exclusive vs Shared Flying Taxi Modal Share Patterns}
\begin{figure}[!htbp]
    \centering
    \includegraphics[width=0.85\textwidth]{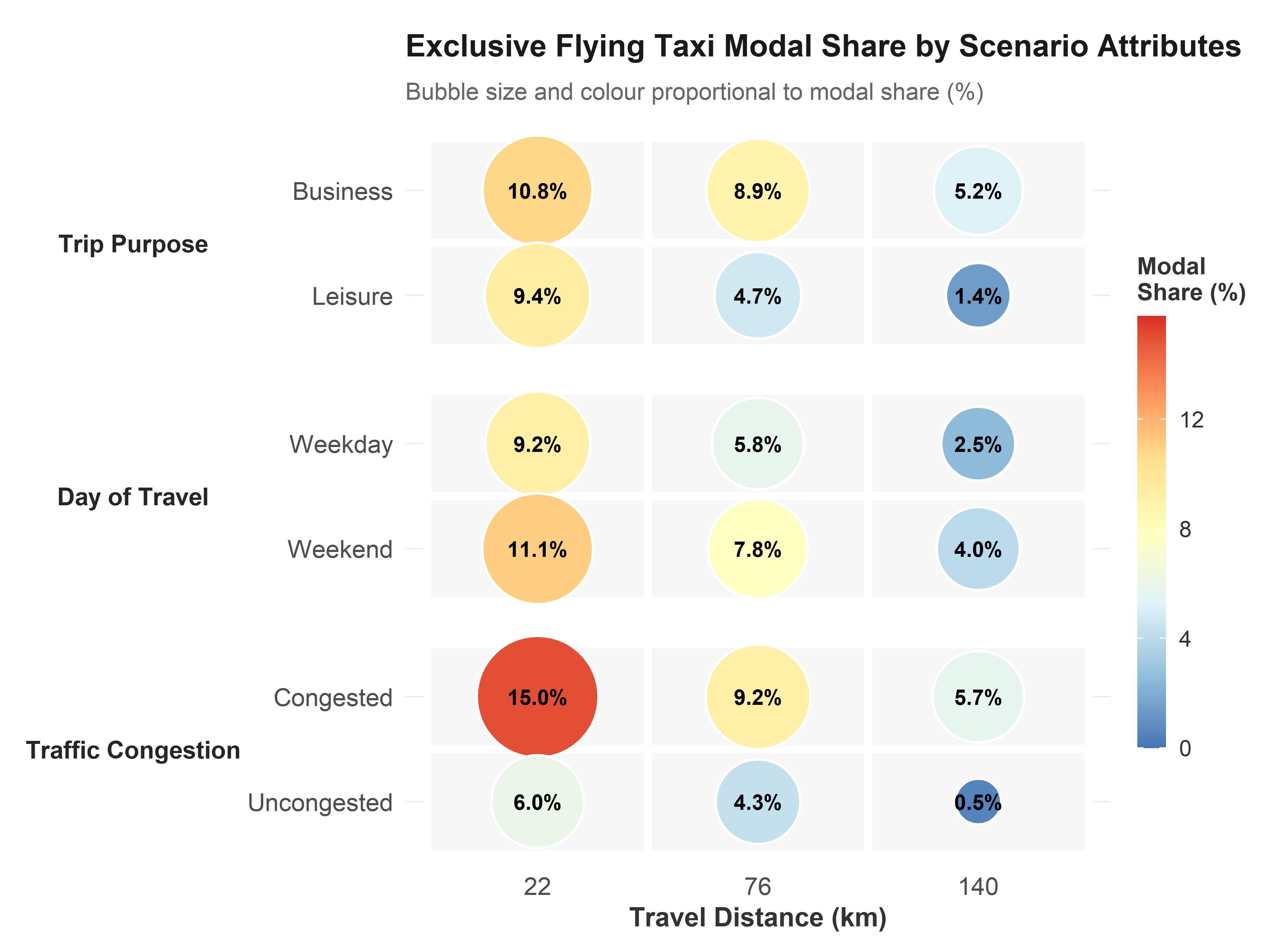}
    \caption{Variation in Exclusive Flying Taxi Modal Share by Scenario Attributes}
    \label{Fig8}
\end{figure}

\begin{figure}[!htbp]
    \centering
    \includegraphics[width=0.85\textwidth]{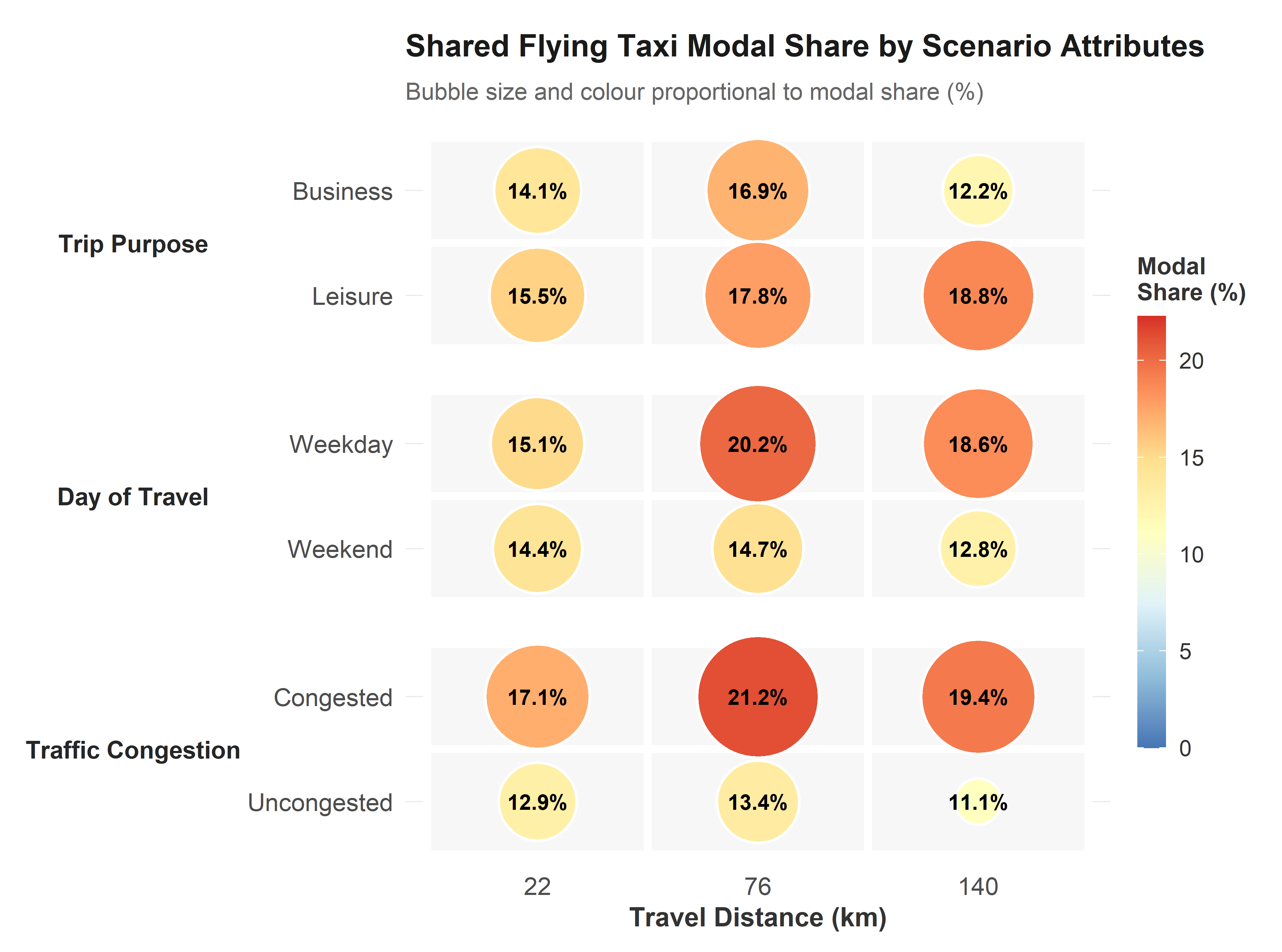}
    \caption{Variation in Shared Flying Taxi Modal Share by Scenario Attributes}
    \label{Fig9}
\end{figure}

Figures~\ref{Fig8} and~\ref{Fig9} provide a detailed comparison of how the modal shares of exclusive and shared flying taxi services vary across the examined scenario attributes. Consistent with the aggregate analysis, both service types exhibit higher modal shares under congested conditions. However, several key differences are evident in respondents’ evaluations of exclusive and shared flying taxi services across varying travel conditions.

A key observation is that shared flying taxis consistently achieve higher modal shares than exclusive flying taxis across all attribute combinations. For instance, under business travel scenarios, the shared flying taxi share ranges from 12.2\% to 16.9\%, compared with 5.2\% to 10.8\% for exclusive services. This difference is also observed across day of travel and traffic congestion attributes, which indicates that shared flying taxis are systematically more attractive than exclusive services, which is likely due to their lower cost, particularly as travel distance increases.

Travel distance also affects the two options differently. Exclusive flying taxi usage is highest for short trips and declines consistently as distance increases. For business trips, the modal share decreases from 10.8\% at 22 km to 8.9\% at 76 km and 5.2\% at 140 km, while for leisure trips it decreases more sharply from 9.4\% to 4.7\% and 1.4\%. In contrast, shared flying taxis reach their highest shares at moderate distances and, under specific conditions, at longer distances. Under weekday conditions, the shared share increases from 15.1\% at 22 km to 20.2\% at 76 km and remains relatively high at 18.6\% at 140 km.

Traffic congestion produces the most pronounced increase in modal share for both services, with a higher effect observed for shared flying taxis. For exclusive flying taxis, congestion raises the modal share from 6.0\% to 15.0\% at 22 km, from 4.3\% to 9.2\% at 76 km, and from 0.5\% to 5.7\% at 140 km. For shared flying taxis, the corresponding increase is from 12.9\% to 17.1\% at 22 km, from 13.4\% to 21.2\% at 76 km, and from 11.1\% to 19.4\% at 140 km. These results indicate that while congestion enhances the attractiveness of both service types, shared flying taxis consistently maintain a higher modal share.

The results also show clear differences in how the two flying taxi options vary by trip purpose. Exclusive flying taxis exhibit higher modal shares for business trips than for leisure trips at all travel distances. In contrast, shared flying taxis exhibit the opposite pattern. Leisure trips are associated with consistently higher modal shares than business trips, increasing from 15.5\% to 17.8\% and 18.8\% across the three distance categories, compared with 14.1\%, 16.9\%, and 12.2\% for business trips. Moreover, when considering the day of travel attribute, shared flying taxis record higher modal shares on weekdays than on weekends across all distances. The difference is particularly pronounced at 76 km and 140 km, where weekday shares reach 20.2\% and 18.6\%, compared with 14.7\% and 12.8\% on weekends. Exclusive flying taxis, by contrast, show higher modal shares on weekends than on weekdays across all distances. For example, at 22 km the exclusive share increases from 9.2\% on weekdays to 11.1\% on weekends, while at 140 km it rises from 2.5\% to 4.0\%. 

The relatively higher share of exclusive flying taxis for business and weekend travel may indicate that respondents associate exclusive services with greater privacy, convenience, and personal control, which become more valuable in these travel contexts. In contrast, shared flying taxis appear to align more closely with cost-sensitive and routine travel situations, particularly under weekday, congested, and moderate-distance conditions where time savings remain important but affordability is also a key consideration. Moreover, while trip purpose and day of travel show limited influence on the overall attractiveness of flying taxi options when considered in aggregate, they play a meaningful role in shaping the relative preference between exclusive and shared operations.

\section{Discussion}
\label{sec:dis}
This section discusses the key findings of the study by interpreting observed mode choice behavior, examining the reasons behind respondents’ choices, and assessing the potential for modal shifts toward flying taxi services. The findings are also discussed in relation to existing literature, and the main contributions of the study are highlighted.

\subsection{Underlying Reasons for Mode Choice Behavior}

In addition to the choice tasks, respondents provided information on the reasons for their selected modes and their willingness to switch to flying taxi options under changing conditions, such as travel time or travel distance. The aggregated results for the first follow-up question across all scenarios are presented in Table~\ref{tab5}. These results offer additional insights into the factors underlying respondents’ mode choice decisions and help explain the modal share patterns observed in the SP experiment. Overall, the findings indicate that respondents’ choices are primarily driven by trade-offs between cost, travel time, convenience, and comfort, with different preferences across transport modes.

\begin{table}[ht]
\caption{Main reasons for selecting transport modes in SP scenarios}
\centering
\begin{tabularx}{\textwidth}{@{}lXc@{}}
\hline
\textbf{Mode} & \textbf{Reason} & \textbf{Percentage (\%)} \\
\hline

\multirow{4}{*}{Ground Taxi} 
& Convenient and easy to book & 51.3 \\
& Comfortable and does not require driving & 30.1 \\
& Offers a balance between speed and price & 11.5 \\
& I am familiar with this mode and trust it & 7.1 \\

\hline

\multirow{4}{*}{Private Car} 
& It is the most cost-effective option & 46.4 \\
& I prefer driving myself & 19.7 \\
& I value flexibility and privacy & 30.1 \\
& I am not comfortable using new or shared transport modes & 3.8 \\

\hline

\multirow{4}{*}{Public Transit} 
& It is the most affordable option & 58.0 \\
& I am environmentally conscious & 15.9 \\
& I am familiar with this option and use it regularly & 9.4 \\
& I avoid driving in traffic whenever possible & 16.7 \\

\hline

\multirow{5}{*}{Exclusive Flying Taxi} 
& It is the fastest option and offers privacy & 44.2 \\
& I want to experience advanced/new technology & 9.3 \\
& I prefer a premium option for business/leisure travel & 37.2 \\
& I prefer to avoid traffic congestion & 8.1 \\
& Flying is more suitable for longer trips / I find air travel enjoyable, even for short distances & 1.2 \\

\hline

\multirow{5}{*}{Shared Flying Taxi} 
& It is the fastest and more affordable than a private flying taxi & 44.6 \\
& I want to experience advanced/new technology & 21.6 \\
& I am open to sharing if it lowers the cost of fast travel & 7.8 \\
& I prefer to avoid traffic congestion & 11.3 \\
& Flying is more suitable for longer trips / I find air travel enjoyable, even for short distances & 14.7 \\

\hline
\end{tabularx}
\label{tab5}
\end{table}

Private vehicles dominate the overall modal share, accounting for the largest proportion of choices across scenarios. This pattern is consistent with respondents’ stated reasons, where cost-effectiveness (46.4\%) and flexibility and privacy (30.1\%) are identified as the main reasons. In addition, 19.7\% of respondents report a preference for driving themselves, which reflects the strong reliance on private vehicles. 

Ground taxi services account for a smaller but notable share of choices (8.7.\%), which is largely explained by their perceived convenience. More than half of the respondents (51.3\%) identify ease of booking as the primary reason for choosing this mode, followed by comfort and the ability to avoid driving (30.1\%). This suggests that ground taxis are primarily valued as a convenient alternative to private vehicles, particularly for users who prefer not to drive while still maintaining a relatively high level of comfort. 

Public transport exhibits a comparatively lower modal share, which is strongly associated with cost considerations. A majority of respondents (58.0\%) select this mode due to its affordability, while smaller shares indicate environmental concerns (15.9\%) and the desire to avoid driving in traffic (16.7\%). These findings highlight that public transport is mainly chosen by cost-sensitive individuals, despite its lower attractiveness in terms of travel time and convenience.

The results for flying taxi services indicate a different pattern of preferences compared to conventional modes. For exclusive flying taxis, respondents are primarily motivated by speed and privacy (44.2\%) and desire for premium travel options (37.2\%). This is consistent with the relatively low overall modal share observed for this option, which suggests that exclusive flaying taxis are perceived as a high-end alternative that appeals to specific user segments, particularly for business or time-sensitive travel. In contrast, shared flying taxis achieve a higher modal share, which can be attributed to their more balanced combination of speed and cost. The main reasons for selecting this option include faster travel at a lower cost compared to exclusive flying taxis (44.6\%) and interest in new technology (21.6\%). In addition, 7.8\% of respondents indicate that they are willing to share the service to reduce travel costs. These findings suggest that shared flying taxis are perceived as a more accessible and practical alternative, particularly for users seeking time savings without incurring the higher costs associated with the exclusive option.

The comparison across modes shows that conventional transport options are mainly chosen based on cost, familiarity, and convenience, while flying taxi services are driven by travel time savings, service quality, and interest in new technology. The difference between exclusive and shared flying taxi services highlights the importance of pricing and service type in shaping choices, with shared services showing greater potential for wider adoption.
 
While the aggregate reasons reported in Table~\ref{tab5} provide an overall picture, the underlying motivations for selecting flying taxi services vary with travel distance. Table~\ref{tab:reasons_by_distance} presents the disaggregated distribution of stated reasons for choosing shared and exclusive flying taxi services across the three distance levels.
 
\begin{table}[ht]
\centering
\caption{Stated reasons for selecting flying taxi services by travel distance}
\label{tab:reasons_by_distance}
\footnotesize
\begin{tabularx}{\textwidth}{@{}Xccc|ccc@{}}
\toprule
& \multicolumn{3}{c|}{\textbf{Shared Flying Taxi}} & \multicolumn{3}{c}{\textbf{Exclusive Flying Taxi}} \\
\cmidrule(lr){2-4} \cmidrule(lr){5-7}
\textbf{Reason} & \textbf{22 km} & \textbf{76 km} & \textbf{140 km} & \textbf{22 km} & \textbf{76 km} & \textbf{140 km} \\
\midrule
Fastest and most affordable / offers privacy & 42.9 & 43.2 & 47.8 & 44.2 & 62.1 & 7.1 \\
Experience new technology & 27.0 & 20.3 & 17.9 & 14.0 & 3.4 & 7.1 \\
Premium option / sharing lowers cost & 6.3 & 6.8 & 10.4 & 34.9 & 24.1 & 71.4 \\
Avoid traffic congestion & 23.8 & 24.3 & 19.4 & 7.0 & 6.9 & 14.3 \\
Suitable for longer trips & 0.0 & 5.4 & 4.5 & 0.0 & 3.4 & 0.0 \\
\addlinespace
\textit{Total responses (n)} & \textit{63} & \textit{74} & \textit{67} & \textit{43} & \textit{29} & \textit{14} \\
\bottomrule
\end{tabularx}
\begin{tablenotes}
\footnotesize
\item Note: Values represent the percentage of respondents within each distance--mode group.
\end{tablenotes}
\end{table}
 
For shared flying taxis, the distribution of stated reasons remains relatively stable across distances, with speed and affordability consistently cited as the primary motivation (42.9\%--47.8\%). However, a notable shift is observed in the relative importance of novelty and cost-sharing motives. The proportion of respondents citing interest in new technology decreases steadily from 27.0\% at 22~km to 17.9\% at 140~km, while the proportion citing willingness to share in order to reduce costs increases from 6.3\% to 10.4\%. This pattern suggests that at shorter distances, shared flying taxis attract more technology-curious users, whereas at longer distances, cost considerations become a stronger motivator. The proportion citing congestion avoidance remains relatively stable across distances (19.4\%--24.3\%), which indicates that this factor is consistently relevant across different trip lengths.
 
The pattern for exclusive flying taxis is different and reveals a pronounced shift in the user profile across distances. At short and medium distances, speed and privacy dominate the stated reasons (44.2\% at 22~km and 62.1\% at 76~km). However, at 140~km, only 7.1\% cite speed and privacy as their primary motivation, while 71.4\% cite a preference for a premium service. This shift is consistent with the sharp decline in exclusive flying taxi modal share at longer distances (from 10.1\% at 22~km to 3.3\% at 140~km), which suggests that the small group of respondents who continue to select exclusive flying taxis at longer distances do so primarily for reasons related to service quality and status rather than travel-time savings. The declining sample size for the exclusive option at 140~km ($n = 14$) warrants caution in interpreting these percentages; nonetheless, the directional shift from time-based to quality-based motivations is clear and consistent with the broader findings of this study.

\subsection{Potential Modal Shift Toward Flying Taxi Services}
\begin{figure}[!htbp]
    \centering
    \includegraphics[width=0.85\textwidth]{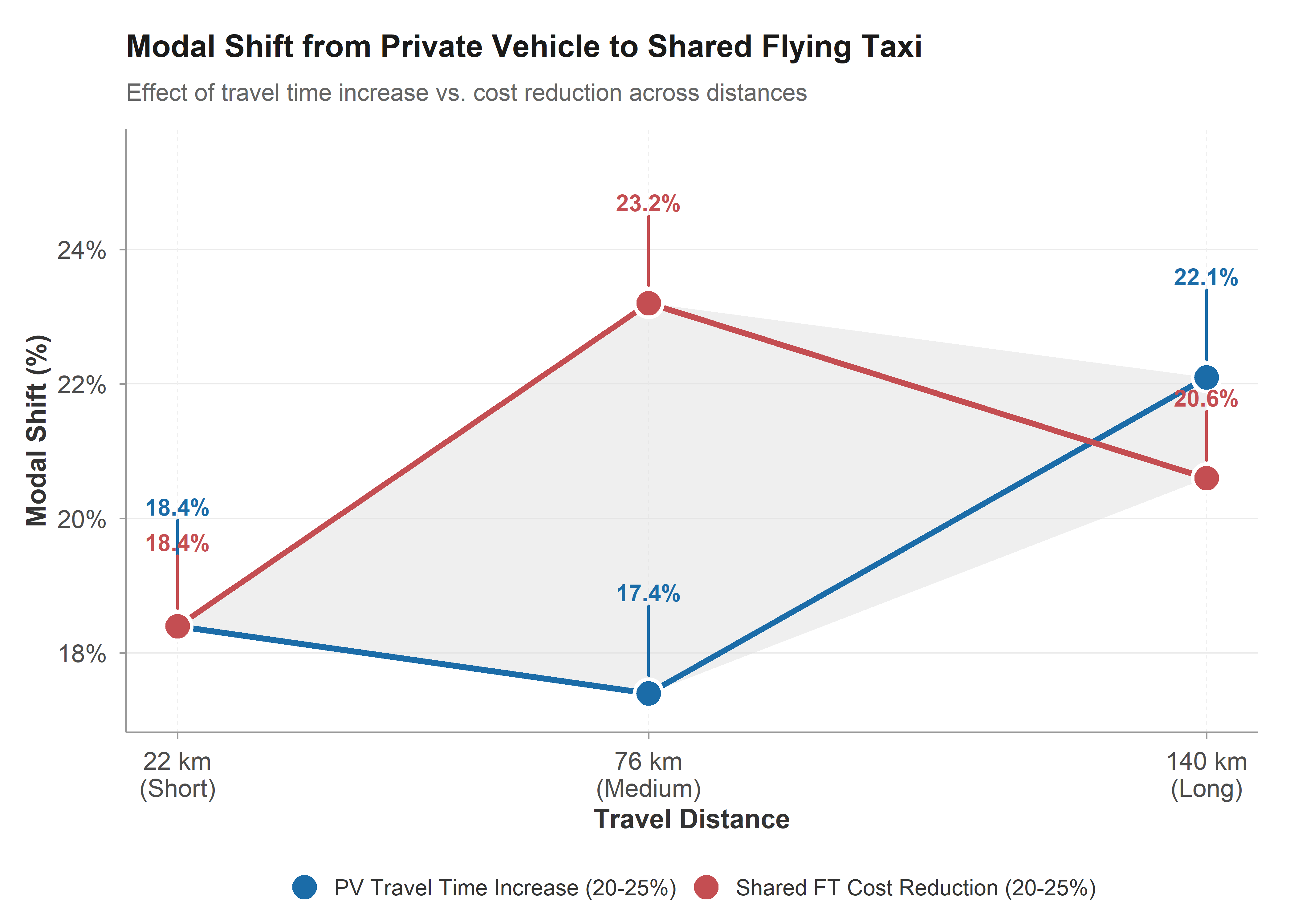}
    \caption{Modal shift intention from private vehicle to shared flying taxi}
    \label{Fig10}
\end{figure}

\begin{figure}[!htbp]
    \centering
    \includegraphics[width=0.85\textwidth]{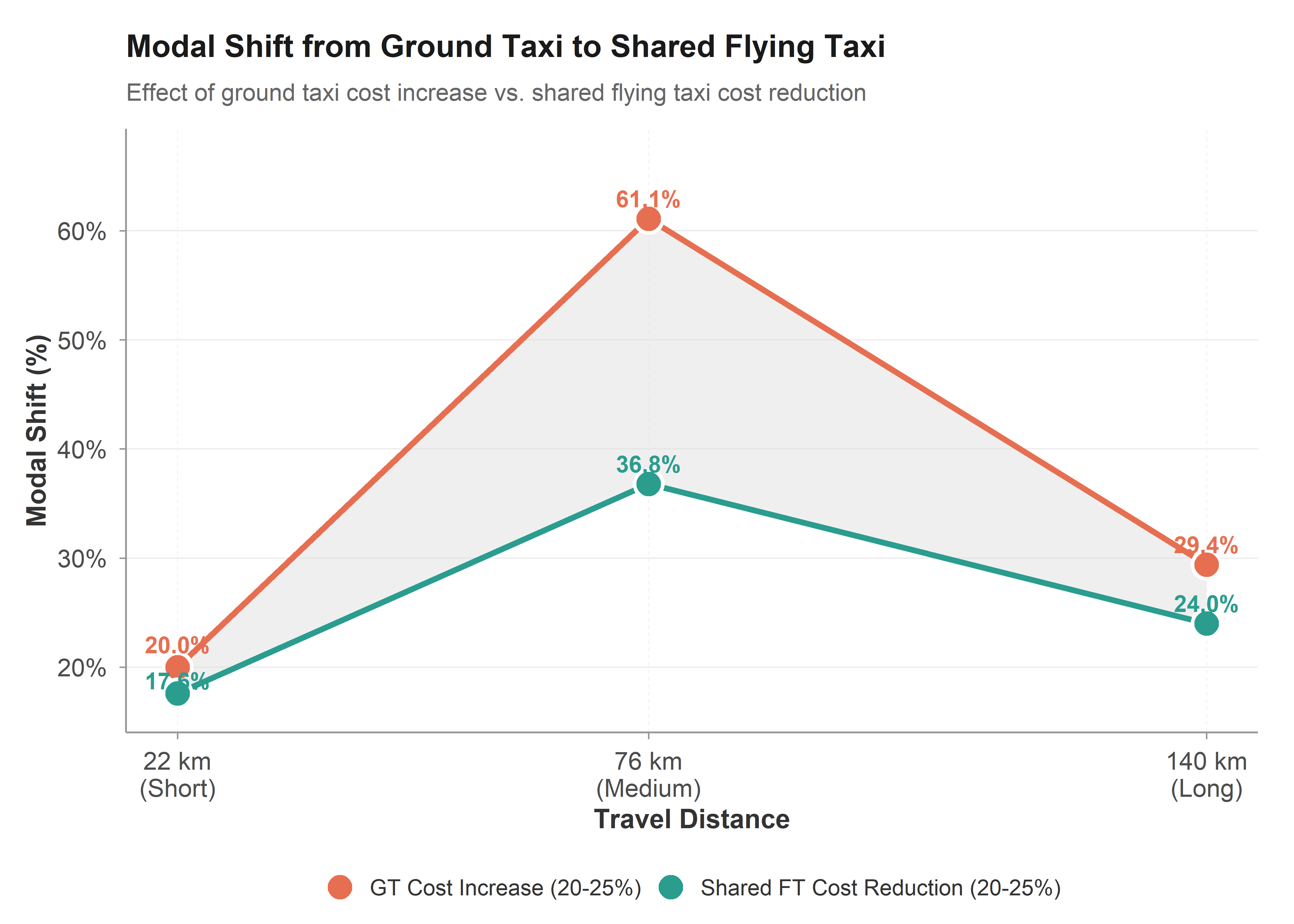}
    \caption{Modal shift intention from ground taxi to shared flying taxi}
    \label{Fig11}
\end{figure}

\begin{figure}[!htbp]
    \centering
    \includegraphics[width=0.85\textwidth]{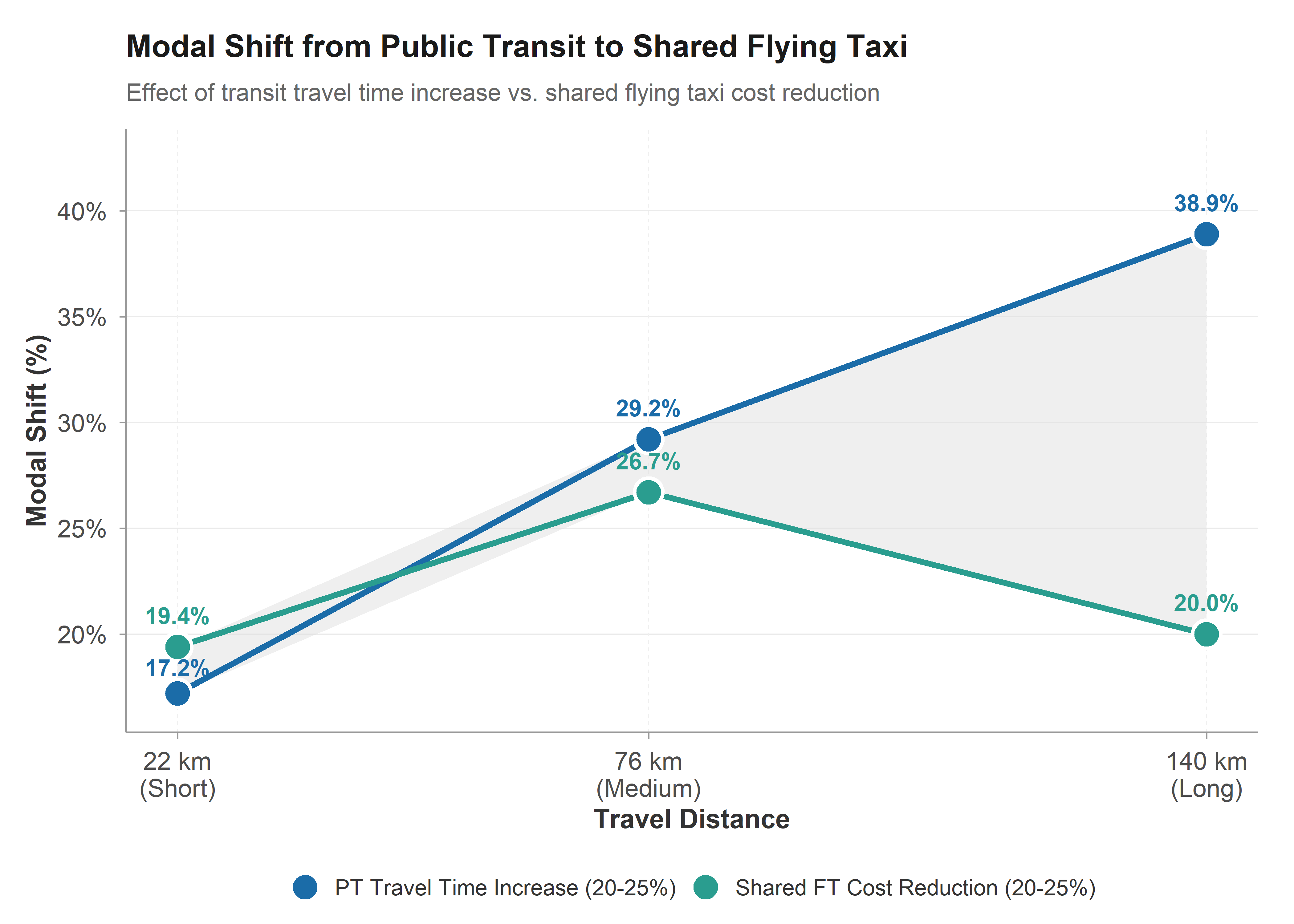}
    \caption{Modal shift intention from public transit to shared flying taxi}
    \label{Fig12}
\end{figure}

\begin{figure}[!htbp]
    \centering
    \includegraphics[width=0.85\textwidth]{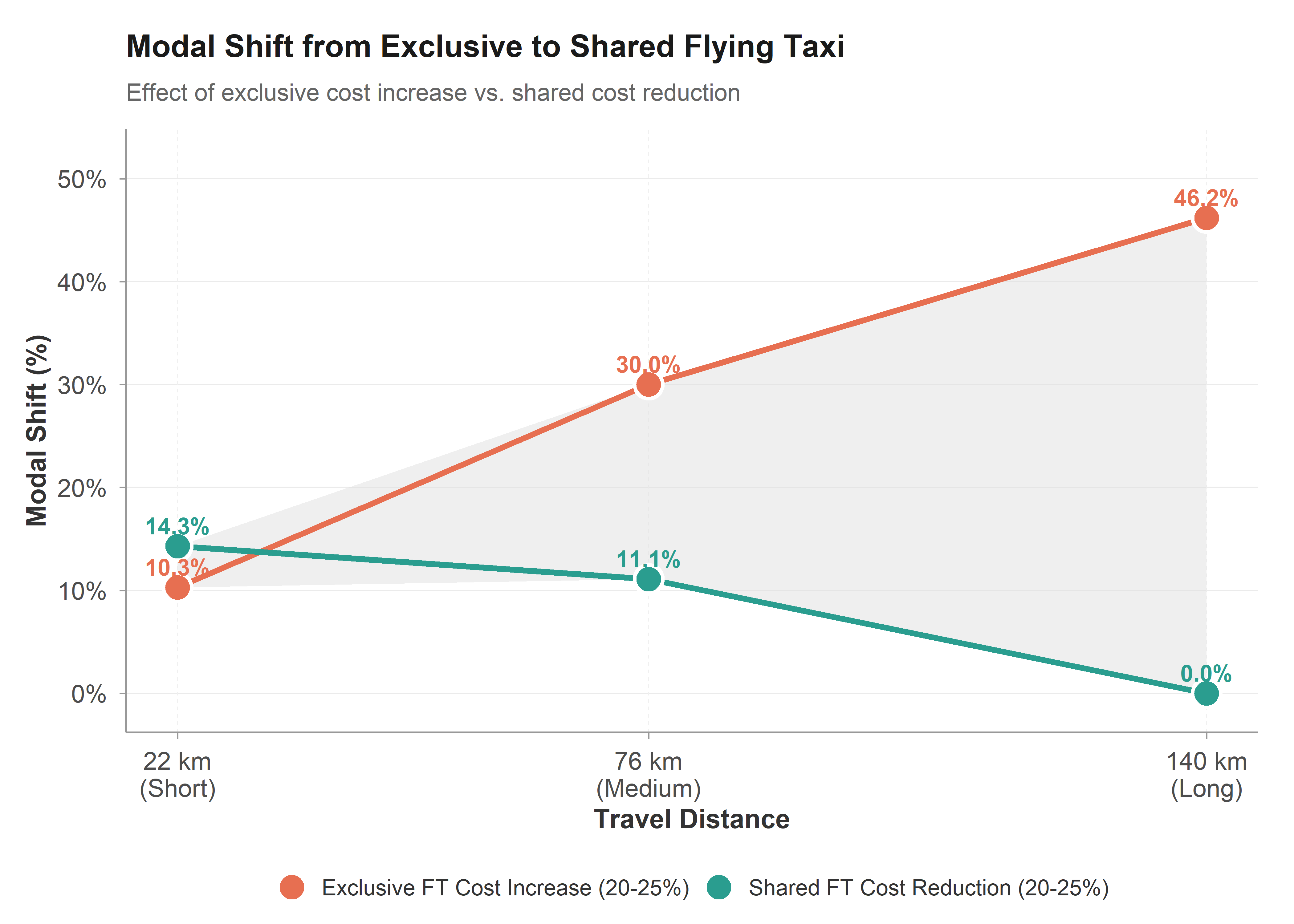}
    \caption{Modal shift intention from exclusive to shared flying taxi}
    \label{Fig13}
\end{figure}

\begin{figure}[!htbp]
    \centering
    \includegraphics[width=0.85\textwidth]{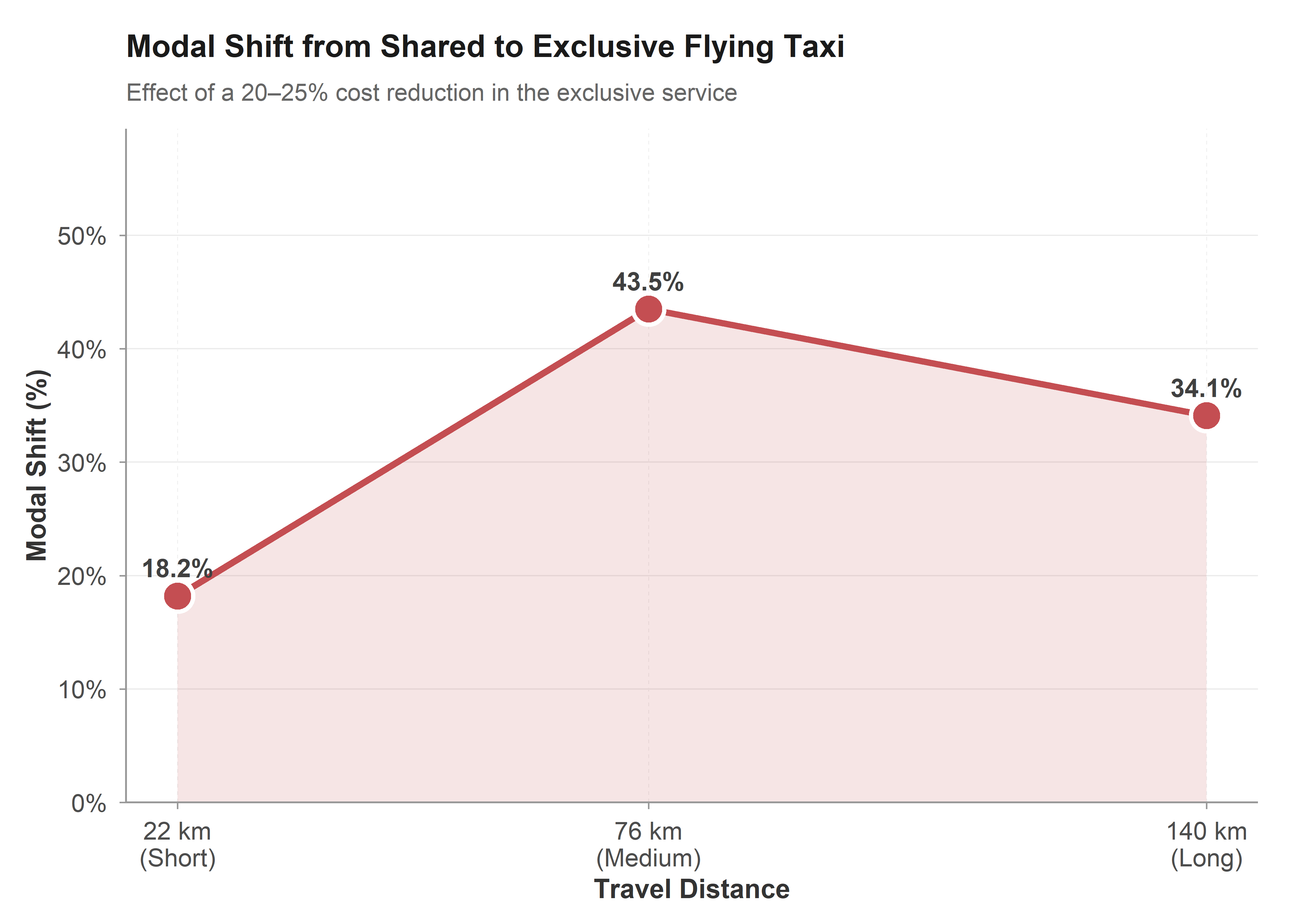}
    \caption{Modal shift intention from shared to exclusive flying taxi}
    \label{Fig14}
\end{figure}

A follow-up question was included after each choice task to assess potential modal shifts, with its formulation depending on the selected mode and the traffic condition in the scenario. For respondents who selected private vehicles or public transport, two types of scenarios were considered. Under congested traffic conditions, respondents were asked whether they would consider switching to a shared flying taxi if their travel time increased by approximately 20–25\%. Under uncongested conditions, they were asked whether they would switch if the cost of shared flying taxi services decreased by a similar margin. For respondents who selected ground taxis or exclusive flying taxis, the scenarios focused on cost changes. Under congested conditions, respondents were asked whether they would consider switching to a shared flying taxi if their travel cost increased by 20–25\%, while under uncongested conditions, they were asked whether they would switch if the cost of shared flying taxi services decreased by a similar margin. Finally, respondents who selected shared flying taxis were asked whether they would consider switching to exclusive flying taxis if its cost decreased by 20–25\%. These scenarios were designed to examine respondents’ sensitivity to travel time and cost, and to provide insights into potential modal shifts under different travel conditions. The results of this analysis are presented in Figures~\ref{Fig10}–\ref{Fig14}.

The results show clear differences in modal shift responses across scenarios, highlighting the relative importance of travel time and cost for each transport mode and across travel distances. For private vehicle users, increasing travel time (by 20–25\%) generally has a stronger effect than reducing the cost of shared flying taxi services. While both scenarios lead to moderate levels of switching, the effect of travel time becomes more pronounced at longer distances (140 km), where modal shift reaches 22.1\%, compared to 20.6\% under cost reduction. This indicates that additional travel time reduces the attractiveness of private vehicles for longer trips; however, the overall shift remains limited, as cost, flexibility, and familiarity were identified as the main reasons for choosing private cars.

In contrast, ground taxi users exhibit the highest responsiveness among all modes. Increasing the cost of ground taxi services leads to substantially higher modal shift compared to reducing the cost of shared flying taxis across all distances, peaking at 61.1\% at 76 km. Although the effect declines at longer distances, it remains higher than the cost reduction scenario, indicating that cost increases act as a stronger push factor than cost reductions as a pull factor. This suggests that ground taxi users are highly sensitive to price changes and represent a key segment for shifting toward shared flying taxi services.

For public transport users, increasing travel time has a consistently stronger effect than reducing the cost of shared flying taxis, particularly as distance increases. Modal shift rises steadily to 38.9\% at 140 km under increased travel time, compared to 20.0\% under cost reduction. This pattern indicates that longer travel times reduce the attractiveness of public transport for longer trips. However, unlike other modes, this shift does not necessarily contribute to sustainability, as it represents a transition from a mass transit mode to a less efficient shared service.

A different pattern is observed for exclusive flying taxi users. Increasing the cost of exclusive flying taxi services has a significantly stronger impact than reducing the cost of shared flying taxis, with modal shift increasing sharply with distance and reaching 46.2\% at 140 km. In contrast, reducing the cost of shared flying taxis has a limited and declining effect, reaching zero at longer distances. This suggests that users of exclusive services are more responsive to increases in their own travel costs than to reductions in shared service costs, likely because privacy and service quality were identified as key reasons for selecting exclusive flying taxis.

Finally, for shared flying taxi users, reducing the cost of exclusive flying taxi services leads to a moderate level of switching, peaking at 43.5\% at 76 km before declining at longer distances. This indicates that while some users are willing to switch to exclusive services when prices decrease, the cost advantage of shared services becomes more important as travel distance increases.

These findings indicate that increasing the cost of existing modes generally produces a stronger modal shift than reducing the cost of shared flying taxi services, particularly for ground taxi and exclusive flying taxi users. Larger travel distances further amplify this effect, as higher baseline travel time and cost reduce the attractiveness of existing modes. From a planning perspective, ground taxi users represent the highest potential for shifting toward shared flying taxi services, followed by exclusive flying taxi users under cost-sensitive conditions. These shifts are particularly important from a sustainability perspective, as they move users toward a shared transport option. However, careful pricing strategies are required to balance the attractiveness of shared services while maintaining a sufficient price difference from exclusive services, in order to avoid unintended shifts from shared to exclusive flying taxi options.

\subsection{Comparison with Existing Literature and Study Contributions}
Before comparing the findings with existing literature, it is useful to assess the consistency between respondents' reported travel behavior and their stated preferences. As shown in Table~\ref{tab:travel_behaviour}, 73.7\% of respondents report using private vehicles as their primary weekday travel mode. In the SP experiment, private vehicles account for 58.1\% of choices across all scenarios. The lower share in the SP context is expected, given that respondents are presented with hypothetical alternatives, including flying taxi services, that are not currently available. Nevertheless, the strong dominance of private vehicles in both the revealed and stated contexts confirms the deeply car-oriented nature of travel behavior in the UAE, which constitutes a significant baseline challenge for the adoption of any emerging mobility service. Similarly, public transport accounts for 10.3\% of reported weekday travel and 10.6\% of SP choices, suggesting that cost-sensitive travelers who rely on public transport maintain a comparable preference in the stated context. Ground taxi and ride-hailing usage shows a moderate increase from 8.5\% (revealed) to 8.7\% (stated), indicating relatively stable demand for this mode. The overall consistency between revealed and stated preferences provides additional support for the validity of the SP results, while the emergence of a 22.6\% combined flying taxi share indicates meaningful potential demand for this new mobility option.

The findings of this study highlight the central role of travel time and cost in shaping mode choice and flying taxi adoption. When considered jointly, flying taxi services account for a notable share of choices across scenarios, indicating their potential as a competitive alternative to conventional modes. Overall, the combined share of flying taxi services increases under congested conditions but decreases with travel distance. However, clear differences are observed between shared and exclusive services across key travel attributes. Shared flying taxis consistently achieve higher modal shares and are more responsive to changes in travel conditions, with their highest shares observed under moderate distances, weekday travel, and congested traffic conditions. In contrast, exclusive flying taxis exhibit lower modal shares and show a decline with increasing travel distance, indicating reduced attractiveness as travel costs increase. Exclusive services are more associated with business travel and weekend trips, suggesting that users value privacy, comfort, and service quality in these contexts. Furthermore, the analysis of modal shift behavior shows that adoption is influenced by both push factors (e.g., increased travel time or cost of existing modes) and pull factors (e.g., reduced cost of shared flying taxis), with cost increases generally having a stronger impact. Overall, these results indicate that while flying taxi services are attractive, their adoption depends strongly on pricing and service design, particularly the availability of cost-effective shared options.

These findings are consistent with previous studies that identify travel time and cost as the most influential factors in flying taxi adoption. For instance, \cite{long2023demand} revealed that trip-related factors, including travel time, cost, and distance are influencing the potential adoption of air taxis. Similarly, \cite{fu2019exploring} and \cite{hwang2023study} found that travel time, cost, and access-related attributes are key determinants of mode choice when flying taxis are compared with conventional transport modes. The importance of cost and time savings observed in this study is also in line with the findings of \cite{karimi2024role}, which highlighted the role of travel cost, time savings, and income in influencing preferences for flying taxis. Overall, the results of this study confirm the key determinants identified in the literature while providing additional insights into how these factors interact across different travel conditions and user groups.

This study contributes to the literature in several important ways. First, it examines traveler preferences under multiple travel conditions, including travel distance, congestion level, day of travel, and trip purpose, while incorporating key service attributes such as travel time and cost. Second, it evaluates flying taxi adoption within a comprehensive mode choice framework that includes private vehicles, public transport, ground taxis, and both exclusive and shared flying taxi services. In particular, the distinction between exclusive and shared flying taxi services provides new insights into how pricing and service design influence user choices. Third, the study integrates both stated mode choice behavior and follow-up responses on reasons for selection and modal shift intentions, allowing for a more comprehensive understanding of travelers’ decision-making processes. Finally, to the best of our knowledge, this study provides the first empirical evidence on traveler preferences for flying taxi services in the UAE.

Based on these findings, several recommendations can be drawn for the planning and implementation of flying taxi services. First, pricing strategies should prioritize the affordability of shared flying taxi services, as cost is a key factor influencing adoption and modal shift. Second, maintaining a clear price differentiation between shared and exclusive services is important to avoid unintended shifts from shared to exclusive options. Third, prioritizing deployment in congested corridors and medium-distance travel markets can enhance the adoption of flying taxi services, where the potential for modal shift is highest. Finally, policies should target user segments with higher sensitivity to cost changes, such as ground taxi users, to encourage a transition toward more sustainable shared mobility options.

\section{Conclusion}
\label{sec:conclusions}

This study examined travelers’ preferences for electric flying taxi services in the UAE under varying travel conditions and service configurations. A stated preference (SP) survey was designed to provide respondents with hypothetical travel scenarios involving multiple transport alternatives, including private vehicles, public transport, ground taxis, and both shared and exclusive flying taxi services. The analysis incorporated key attributes such as travel time and cost, along with contextual factors including travel distance, congestion conditions, day of travel, and trip purpose. In addition, follow-up questions were used to capture the underlying reasons for mode choice and to assess potential modal shifts under changes in travel conditions.

The results highlight the importance of travel time and cost for mode choice and flying taxi adoption. Overall, flying taxi services account for a notable share of choices, with higher shares observed under congested conditions and lower shares at longer travel distances, while trip purpose and day of travel have a limited impact on overall adoption. However, these two factors play an important role in shaping the relative preference between shared and exclusive services. Clear differences are observed between the two service types. Although both shared and exclusive flying taxis exhibit higher modal shares under congested conditions, shared services consistently achieve higher shares across all scenarios. Shared flying taxis are more attractive at moderate distances, during weekdays, and for leisure trips, whereas exclusive services are more associated with shorter distances, business travel, and weekend travel. The modal shift analysis further shows that increases in the cost or travel time of existing modes have a stronger effect than reductions in shared flying taxi costs. Overall, the findings emphasize the importance of pricing and service design, with shared services offering greater potential for wider adoption.

The main limitations of this study include the limited representation of respondents from the Fujairah and Umm Al Quwain emirates, which may affect the generalization of the results. In addition, the study adopts a descriptive and exploratory approach and does not incorporate behavioral modeling to statistically quantify the effects of travel attributes or capture preference heterogeneity. Future research should address these limitations by improving geographic coverage and developing discrete choice models to better understand individual preferences and quantify the impacts of various factors, including socio-economic characteristics, travel behavior, attitudinal factors, and scenario attributes, on preferences for shared and exclusive flying taxi services.

\section*{Authorship contribution statement}
The authors confirm their contribution to the paper as follows: \textbf{Nael Alsaleh}.: Conceptualization, Methodology, Data curation, Investigation, Formal analysis, Software, Visualization, Writing - original draft, and Writing - review \& editing, Supervision, Project administration, Funding acquisition. \textbf{Tareq Alsaleh}.: Conceptualization, Methodology, Formal analysis, Software, Visualization, Writing - original draft, and Writing - review \& editing. \textbf{Fayez Moutassem}.: Formal analysis, Visualization, Writing - original draft. \textbf{Noura Falis}.: Formal analysis, Visualization, Writing - original draft. \textbf{Zainab Islam}.:  Formal analysis, Visualization, Writing - original draft.

\section*{Acknowledgements}
This research was supported by a seed grant (ENGR/008/26) provided by the Office of Research and Sustainability at the American University of Ras Al Khaimah (AURAK), UAE. The authors would like to thank AURAK University for supporting the research, and all survey respondents for their time and participation.

\section*{Declaration of generative AI use}
The AI tool (ChatGPT) was used only for grammar and language refinement, with no AI-generated original content. The authors reviewed and edited the content as needed and take full responsibility for the content of the published article.

\section*{Conflict of Interest}
The authors declare that they have no known competing financial interests or personal relationships that could have appeared to influence the work reported in this paper.

\newpage

\appendix
\setcounter{table}{0}
\renewcommand{\thetable}{A\arabic{table}}
\begin{landscape}
\section{Details of SP Scenarios}
\label{app:scenarios}

Table~\ref{tab:scenarios} provides travel time (minutes) and cost (AED) values used in the stated preference scenarios. 

\begin{table}[H]
\centering
\caption{Travel time (minutes) and cost (AED) values used in the stated preference scenarios}
\label{tab:scenarios}

\adjustbox{max width=\linewidth}{
\begin{tabular}{ccccccccc ccccccc}
\toprule
& & & & & \multicolumn{2}{c}{Ground Taxi} & \multicolumn{2}{c}{Private Vehicle} & \multicolumn{2}{c}{Public Transit} & \multicolumn{2}{c}{Exclusive Flying Taxi} & \multicolumn{2}{c}{Shared Flying Taxi} \\
\cmidrule(lr){6-7} \cmidrule(lr){8-9} \cmidrule(lr){10-11} \cmidrule(lr){12-13} \cmidrule(lr){14-15}

Scenario & Distance & Purpose & Day & Traffic 
& Time & Cost 
& Time & Cost 
& Time & Cost 
& Time & Cost 
& Time & Cost \\

\midrule

1 & 140 & Business & Weekday & Congested & 95 & 360 & 95 & 90 & 165 & 40 & 35 & 1150 & 35 & 575 \\
2 & 140 & Business & Weekday & Uncongested & 75 & 345 & 75 & 70 & 145 & 40 & 35 & 1150 & 35 & 575 \\
3 & 140 & Business & Weekend & Congested & 100 & 365 & 100 & 95 & 165 & 40 & 35 & 1150 & 35 & 575 \\
4 & 140 & Business & Weekend & Uncongested & 75 & 345 & 75 & 70 & 145 & 40 & 35 & 1150 & 35 & 575 \\
5 & 140 & Leisure & Weekday & Congested & 95 & 360 & 95 & 90 & 165 & 40 & 35 & 1150 & 35 & 575 \\
6 & 140 & Leisure & Weekday & Uncongested & 75 & 345 & 75 & 70 & 145 & 40 & 35 & 1150 & 35 & 575 \\
7 & 140 & Leisure & Weekend & Congested & 100 & 365 & 100 & 95 & 165 & 40 & 35 & 1150 & 35 & 575 \\
8 & 140 & Leisure & Weekend & Uncongested & 75 & 345 & 75 & 70 & 145 & 40 & 35 & 1150 & 35 & 575 \\

9 & 76 & Business & Weekday & Congested & 65 & 210 & 65 & 45 & 125 & 25 & 20 & 600 & 20 & 300 \\
10 & 76 & Business & Weekday & Uncongested & 55 & 195 & 55 & 35 & 115 & 25 & 20 & 600 & 20 & 300 \\
11 & 76 & Business & Weekend & Congested & 70 & 215 & 70 & 50 & 125 & 25 & 20 & 600 & 20 & 300 \\
12 & 76 & Business & Weekend & Uncongested & 55 & 195 & 54 & 35 & 115 & 25 & 20 & 600 & 20 & 300 \\
13 & 76 & Leisure & Weekday & Congested & 65 & 210 & 65 & 45 & 125 & 25 & 20 & 600 & 20 & 300 \\
14 & 76 & Leisure & Weekday & Uncongested & 55 & 195 & 55 & 35 & 115 & 25 & 20 & 600 & 20 & 300 \\
15 & 76 & Leisure & Weekend & Congested & 70 & 215 & 70 & 50 & 125 & 25 & 20 & 600 & 20 & 300 \\
16 & 76 & Leisure & Weekend & Uncongested & 55 & 195 & 54 & 35 & 115 & 25 & 20 & 600 & 20 & 300 \\

17 & 22 & Business & Weekday & Congested & 35 & 80 & 35 & 25 & 55 & 7.5 & 8 & 350 & 8 & 175 \\
18 & 22 & Business & Weekday & Uncongested & 18 & 60 & 18 & 15 & 49 & 7.5 & 8 & 350 & 8 & 175 \\
19 & 22 & Business & Weekend & Congested & 40 & 85 & 40 & 30 & 49 & 7.5 & 8 & 350 & 8 & 175 \\
20 & 22 & Business & Weekend & Uncongested & 18 & 60 & 18 & 15 & 49 & 7.5 & 8 & 350 & 8 & 175 \\
21 & 22 & Leisure & Weekday & Congested & 35 & 80 & 35 & 25 & 55 & 7.5 & 8 & 350 & 8 & 175 \\
22 & 22 & Leisure & Weekday & Uncongested & 18 & 60 & 18 & 15 & 49 & 7.5 & 8 & 350 & 8 & 175 \\
23 & 22 & Leisure & Weekend & Congested & 40 & 85 & 40 & 30 & 49 & 7.5 & 8 & 350 & 8 & 175 \\
24 & 22 & Leisure & Weekend & Uncongested & 18 & 60 & 18 & 15 & 49 & 7.5 & 8 & 350 & 8 & 175 \\

\bottomrule
\end{tabular}
}

\end{table}

\section{Mode Choice Outcomes for SP Scenarios}

Table~\ref{tab:mode_choice_distribution} presents the observed distribution of mode choices across the SP scenarios.

\begin{table}[H]
\centering
\caption{Observed mode choice distribution across SP scenarios}
\label{tab:mode_choice_distribution}

\adjustbox{max width=\linewidth}{
\begin{tabular}{ccccc cccccc}
\toprule
& \multicolumn{4}{c}{Attributes} & \multicolumn{5}{c}{Mode Choice Distribution} & \\
\cmidrule(lr){2-5} \cmidrule(lr){6-10}

Scenario & Travel Distance & Trip Purpose & Day of Travel & Traffic Condition 
& Ground Taxi & Private Vehicle & Public Transit & Exclusive Flying Taxi & Shared Flying Taxi 
& Total Responses \\

\midrule

1  & 140 & Business & Weekday & Congested   & 4 & 30 & 3 & 5 & 8  & 50 \\
2  & 140 & Business & Weekday & Uncongested & 3 & 37 & 4 & 0 & 5  & 49 \\
3  & 140 & Business & Weekend & Congested   & 4 & 36 & 6 & 6 & 10 & 62 \\
4  & 140 & Business & Weekend & Uncongested & 8 & 35 & 6 & 0 & 3  & 52 \\
5  & 140 & Leisure  & Weekday & Congested   & 6 & 33 & 3 & 0 & 17 & 59 \\
6  & 140 & Leisure  & Weekday & Uncongested & 7 & 25 & 2 & 0 & 7  & 41 \\
7  & 140 & Leisure  & Weekend & Congested   & 3 & 37 & 5 & 2 & 9  & 56 \\
8  & 140 & Leisure  & Weekend & Uncongested & 7 & 34 & 8 & 1 & 7  & 57 \\

9  & 76 & Business & Weekday & Congested   & 4 & 22 & 4 & 7 & 13 & 50 \\
10 & 76 & Business & Weekday & Uncongested & 7 & 41 & 2 & 1 & 9  & 60 \\
11 & 76 & Business & Weekend & Congested   & 2 & 26 & 8 & 7 & 8  & 51 \\
12 & 76 & Business & Weekend & Uncongested & 3 & 34 & 5 & 4 & 6  & 52 \\
13 & 76 & Leisure  & Weekday & Congested   & 7 & 27 & 5 & 3 & 10 & 52 \\
14 & 76 & Leisure  & Weekday & Uncongested & 6 & 24 & 5 & 1 & 10 & 46 \\
15 & 76 & Leisure  & Weekend & Congested   & 5 & 34 & 7 & 3 & 15 & 64 \\
16 & 76 & Leisure  & Weekend & Uncongested & 3 & 39 & 3 & 3 & 3  & 51 \\

17 & 22 & Business & Weekday & Congested   & 4 & 18 & 8 & 8 & 6  & 44 \\
18 & 22 & Business & Weekday & Uncongested & 4 & 32 & 8 & 2 & 11 & 57 \\
19 & 22 & Business & Weekend & Congested   & 4 & 20 & 8 & 8 & 6  & 46 \\
20 & 22 & Business & Weekend & Uncongested & 3 & 45 & 6 & 5 & 7  & 66 \\
21 & 22 & Leisure  & Weekday & Congested   & 4 & 30 & 6 & 7 & 10 & 57 \\
22 & 22 & Leisure  & Weekday & Uncongested & 7 & 33 & 11 & 3 & 6  & 60 \\
23 & 22 & Leisure  & Weekend & Congested   & 3 & 19 & 7 & 6 & 11 & 46 \\
24 & 22 & Leisure  & Weekend & Uncongested & 3 & 31 & 6 & 4 & 6  & 50 \\

\bottomrule
\end{tabular}
}

\end{table}

\end{landscape}

\bibliographystyle{elsarticle-harv}
\singlespacing
\bibliography{References.bib}
\end{document}